\documentclass[12pt]{article}
\usepackage[dvips]{graphicx,color}
\usepackage{rotating}

\usepackage{amssymb}

\def\mdmatm{\Delta m^2_{32}}
\def\dmatm{$\mdmatm$}
\def\mdmsol{\Delta m^2_{21}}
\def\dmsol{$\mdmsol$}
\def\numunue{$\nu_\mu \rightarrow \nu_e$}
\def\anumunue{$\bar\nu_\mu \rightarrow \bar\nu_e$}
\def\meV{e\mbox{V}}
\def\eV{$\meV$}
\def\MeV{M\eV}
\def\GeV{G\eV}

\begin{document}

\begin{titlepage}

  \begin{center}
    
    \noindent{\large \bf 
      Neutrino Oscillation 
      Experiments for Precise Measurements of Oscillation 
      Parameters and Search for \numunue{} Appearance 
      and CP Violation. 

      LETTER OF INTENT to Brookhaven National Laboratory.

    }
  \end{center}

\begin{center}
 D. Beavis, M. Diwan, R. Fernow, 
 J. Gallardo,
 S. Kahn, H. Kirk, W. Marciano, 
W. Morse,  
Z. Parsa, R. Palmer, T. Roser, 
N. Samios, Y. Semertzidis, 
N. Simos,  B. Viren, W. Weng\\
{\sl
Brookhaven National Laboratory
Box 5000, Upton, NY 11973-5000 }
\smallskip

 W. Frati, J.R. Klein, K. Lande, A.K. Mann, 
R. Van Berg and P. Wildenhain \\
{\sl 
University of Pennsylvania
Philadelphia, PA 19104-6396 }

\smallskip

R. Corey\\
{\sl South Dakota School of Mines and Technology
Rapid City, S.D. 57701}

\smallskip

D.B.~Cline, K.~Lee, B.~Lisowski, P.F.~Smith \\
{\sl Department of Physics and Astronomy, University of California, Los Angeles, CA 90095 USA}

\smallskip

A.~Badertscher, A.~Bueno, L.~Knecht, G.~Natterer, S.~Navas, A.~Rubbia  \\
{\sl Institut für Teilchenphysik, ETHZ, CH-8093 Z\"urich, Switzerland}

\smallskip

R.F.~Burkart, W.~Burgett, E.J.~Feynves \\
{\sl Department of Physics, University of Texas at Dallas, 
Richardson, TX 75083 USA} 

\smallskip

J.G.~Learned \\
{\sl Department of Physics and Astronomy, University of Hawaii, 
Honolulu, HI 96822 USA}

\noindent
V. Palladino \\
{\sl Universita di Napoli ``Federico II", 80138 Napoli, Italy}

\smallskip

I.~Mocioiu, R.~Shrock \\ 
{\sl C.N.~Yang Institute for Theoretical Physics, 
State University of New York, Stony Brook, NY 11974 USA}

\smallskip 

C.~Lu, K.T.~McDonald \\
{\sl Joseph Henry Laboratories, Princeton University, 
Princeton, NJ 08544 USA}

April 23, 2002

\end{center}

\vspace{.2in}

\end{titlepage}

BLANK PAGE  

\newpage 

\begin{abstract}

  The possibility of making a low cost, very intense   high
  energy proton source  at the Brookhaven Alternating Gradient
Synchrotron  (AGS) 
  along with  the forthcoming
  new large underground detectors at either  the National Underground
  Science Laboratory (NUSL) in Homestake, South Dakota or at the Waste
  Isolation Pilot Plant (WIPP) in Carlsbad, New Mexico, allows us to
  propose a program of experiments that will address fundamental
  aspects of neutrino oscillations and CP-invariance violation.  This
  program of experiments is unique because of the extra-long
   baseline of more than 2500 km from Brookhaven National
  Laboratory to the underground laboratories in the West, the high
  intensity of the proposed conventional neutrino beam, and the
  possibility of constructing a very large array of water Cerenkov
  detectors with total mass approaching 1 Megaton.  As part of this
  program we also  consider experiments at moderately long
  baselines ($\sim$400~km) using other detector technologies that can
  yield valuable and complementary information on neutrino
  oscillations.  This letter of intent focuses on the design and
  construction of the necessary AGS upgrades and the new neutrino beam
  which will initially have a proton beam of power $\sim$0.5~MW; the
  power will then be  upgraded to $\sim$1.3 MW in a later phase.

\end{abstract}



\newpage


\section{ Introduction}

\vspace{1ex}

This is a letter of intent to build a new high intensity neutrino beam
at BNL in two phases.  The first phase will involve AGS upgrades
consisting of an addition to the present LINAC to yield a total LINAC
energy of 400 \MeV, and an accumulator ring with permanent magnets
mounted in the AGS tunnel.  This will allow  injection into the AGS
at 2.5 \GeV
and an AGS cycle time of 1 Hz.  At 28 \GeV, the
intensity should be $1.2 \times 10^{14}$ protons per pulse and the
power 0.53 MW.  The second phase will require new power supplies for
the AGS and booster magnets for rapid cycling, an RF upgrade, and
additional shielding; these will produce an AGS cycle time of 2.5 Hz
and 1.3 MW power.  The new intense proton beam will be used to produce
a conventional horn focussed neutrino beam directed at the far detectors.
We propose to send the beam to two different distances: the very long
distance of 2540 km (2880 km) to the Homestake (WIPP) 
\cite{homestk, wipp}  laboratory, and
a much shorter distance of 400 km 
to a new detector location 
in upstate New York.

It is now well known that the strongest evidence for neutrino
oscillations so far comes from astrophysical observations of
atmospheric neutrinos with $\mdmatm = (1.6 - 4.0) \times 10^{-3}
\meV^2$ and maximal mixing \cite{sk}
and from 
solar neutrinos with $\mdmsol \sim (2
-10) \times 10^{-5} \meV^2$ and the LMA solution for solar
neutrinos~\cite{sno}.  
The observation by the LSND
experiment~\cite{lsnd} will soon be re-tested at Fermilab by the
mini-Boone \cite{boon} experiment, therefore we will not discuss it 
further in this document.
There are several accelerator based experiments (K2K,
MINOS, and CNGS)  \cite{k2k, j2k, minos, cngs} currently  
 in construction phase or taking data 
 to confirm the atmospheric neutrino signatures for
oscillations. 
 There is now a consensus that there are four main goals
in the field of neutrino oscillations 
 that should be addressed soon with accelerator neutrino
beams:

\begin{enumerate}
\item Precise determination of \dmatm{} and definitive observation of
  oscillatory behavior.
\item Detection of \numunue{} in the appearance mode.  If the measured
  $\Delta m^2$ for this measurement is near \dmatm{} then this
  appearance signal will show that $\left|U_{e3}\right|^2 (=
  \sin^2\theta_{13})$ from the neutrino mixing matrix in the standard
  parameterization is non-zero.
\item Detection of the matter enhancement effect in \numunue{} in the
  appearance mode. This effect will also allow us to measure the sign
  of \dmatm{}; i.e. which neutrino is heavier.
\item Detection of CP violation in neutrino physics.  The neutrino
  CP-violation in Standard Model neutrino physics comes from the phase
  multiplying $\sin \theta_{13}$ in the mixing matrix. This can be
  detected by observing an asymmetry in the oscillation rates
  \numunue{} versus \anumunue{}.

\end{enumerate}

In the following we will briefly describe how all of these goals can
be achieved under reasonable assumptions for the various parameters
using the new intense AGS based beam and the long and very long
baselines.

In the first section of this letter of intent we describe the physics
purpose of the new beam sent along the baseline of 2540 km (results for
2880 km are similar).  The second section describes the physics case
for the shorter distance of 400 km.  The third section briefly
discusses the design and construction of the AGS upgrades.  And the
last section describes the construction of the new neutrino beam.

\section{Very Long Baseline Experiment} 

\vspace{1ex}

\begin{figure}
  \begin{center}
  \includegraphics*[width=\textwidth]{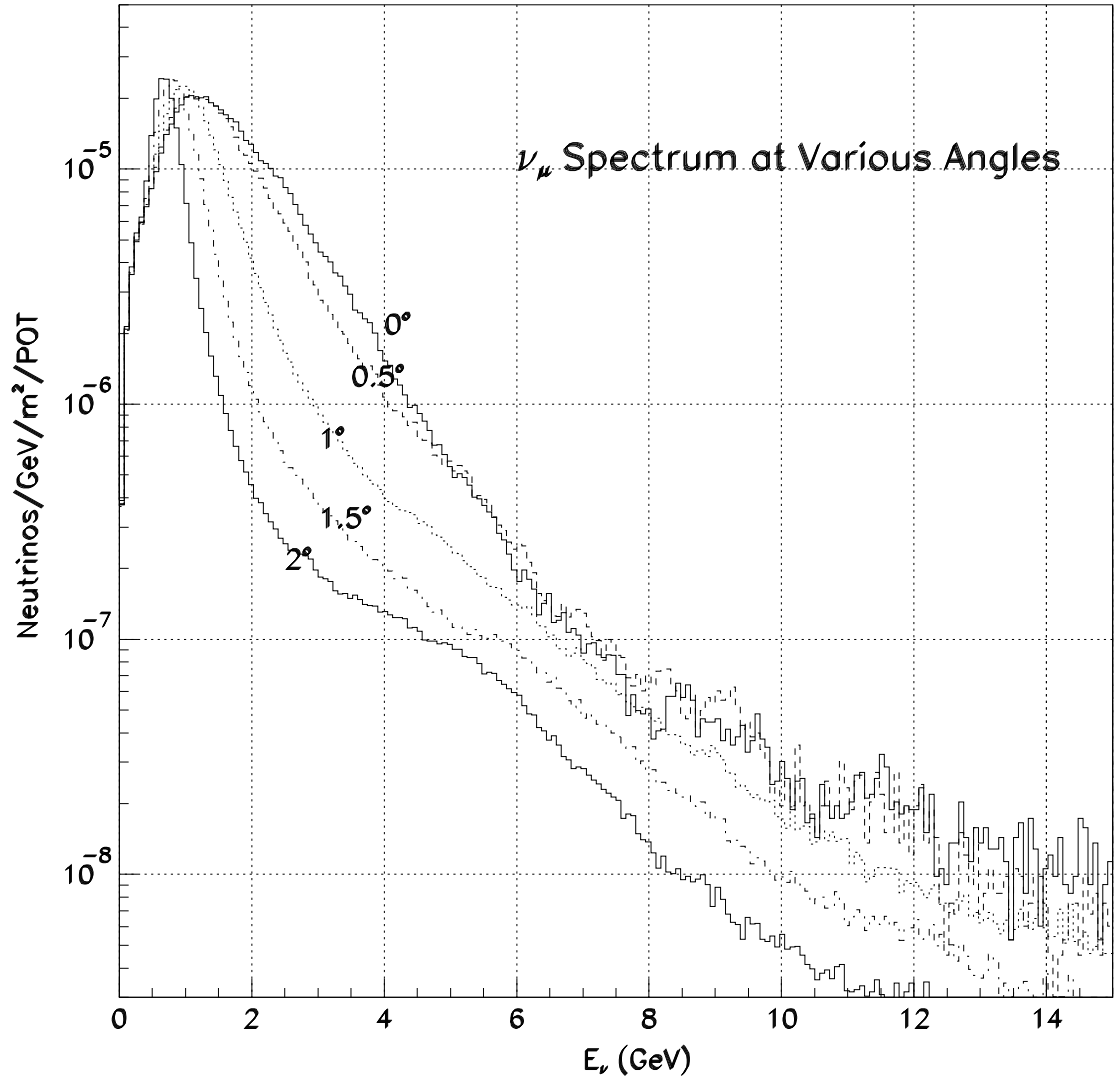}
  \caption{Wide band horn focussed neutrino spectrum for 28 \GeV{} protons. 
    Spectrum of neutrinos are calculated at various angles with respect to the
    200 m decay tunnel axis at the AGS and at a distance of 1 km from the target.} 
  \label{bnlspec}
  \end{center}
\end{figure}

\begin{figure}[htbp]
  \begin{center}
  \includegraphics*[width=\textwidth]{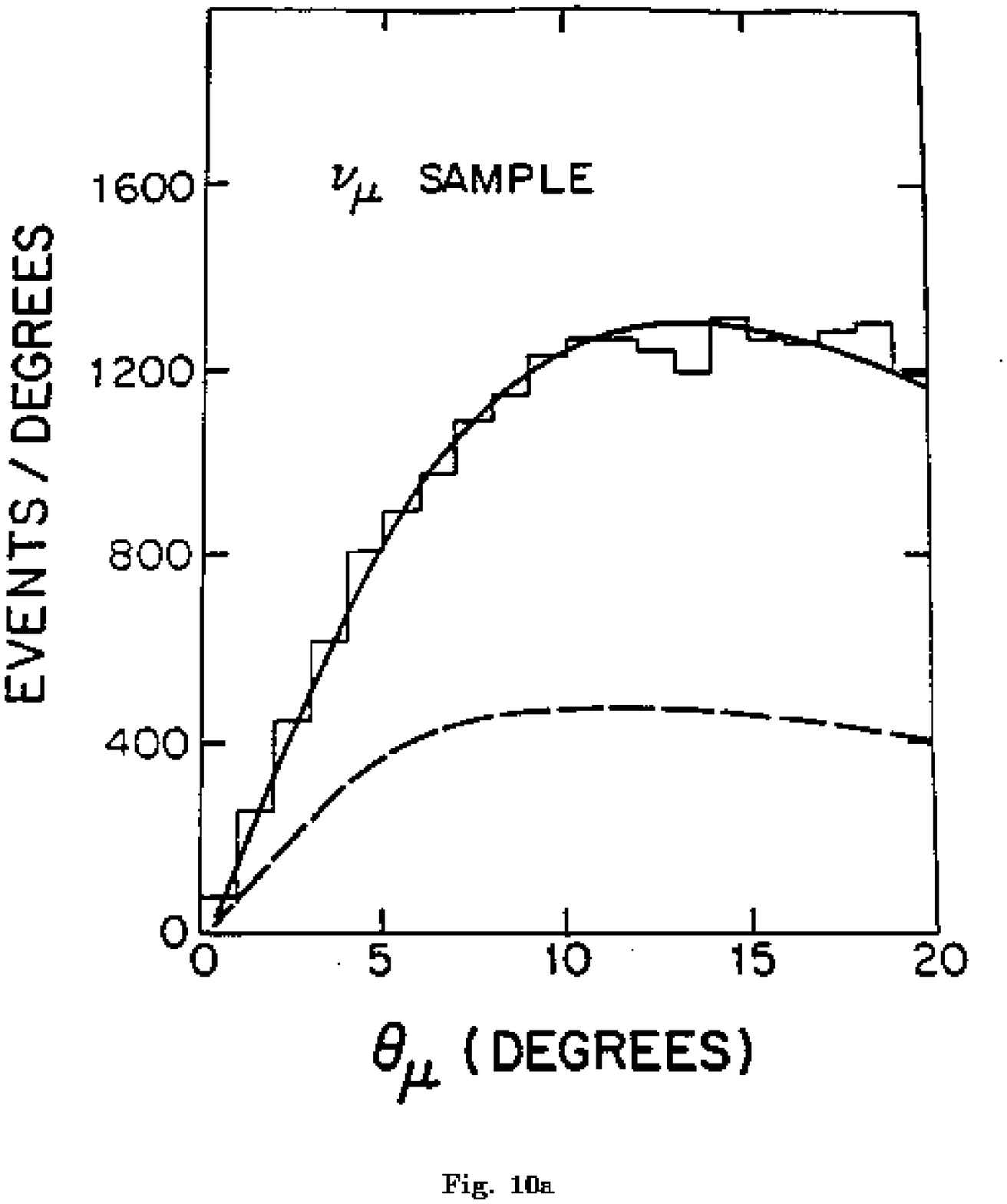}
  \caption{
Angular distribution of muons from the process 
 $\nu_\mu n \rightarrow \mu^- p$ (top curve) and 
background from  
 $\nu_\mu N \rightarrow \mu^- N' \pi$ (bottom curve).
The histogram is data from E734 and lines are Monte Carlo.}
  \label{fig:e734mu}
  
  \end{center}
\end{figure}

The calculated energy distributions of a $\nu_{\mu}$ beam produced by
28 \GeV{} protons is shown in Fig.~\ref{bnlspec}~\cite{e889}.  The
$0^\circ$ calculation has been shown consistent with neutrino beam
data~\cite{e734}.  The spectrum peaks at about 1 \GeV{} with a total
spread at half intensity of about 1 \GeV{}.  Further work on the
optimization of this spectrum for the very long baseline experiment is
on going.  We calculate the event rate without oscillations assuming a
0.5 MW proton beam power with 28 \GeV{} protons ($1.1 \times 10^{14}$
ppp), a 0.5 MT fiducial mass detector and 5 years of running.  Because
the AGS can run in a parasitic mode to RHIC we expect to get beam for
as much as $1.8\times 10^7$ sec per year.  However, we conservatively
assume only $1.0\times 10^7$ sec of AGS running 
per year here.  Using these
parameters and the relevant cross section, we calculate that the
number of quasi-elastic charged current muon neutrino events in a
detector located at 2540 km will be $\sim 5700$ in five years running.
The number of neutrino events of all types will be approximately twice
this number.  This large statistics combined with the long baseline
makes many of the following important measurements possible.

The angular distribution of the muons from the quasi-elastic process
$\nu_{\mu} + n \rightarrow \mu^- + p$ produced by the $0^o$ beam in
Fig.~\ref{bnlspec} are shown in Fig.~\ref{fig:e734mu}; the principal
background, $\nu_{\mu} + n \rightarrow \mu^- + p + \pi$ is also shown
\cite{e734d}.
A variety of strategies are possible to reduce this background further
in a water Cerenkov detector.
 Knowing the direction of
an incident $\nu_{\mu}$ accurately and measuring the angle of the
observed muon allows the energy of the $\nu_{\mu}$ to be calculated,
up to Fermi momentum effects. 
This method is used by the currently running K2K experiment
\cite{k2k}.  The known capability of large water Cerenkov detectors
indicates that at energies lower than 1 \GeV{} the $\nu_\mu$ energy
resolution will be dominated by Fermi motion and 
nuclear effects\cite{kasuga}.
  The
contribution to the resolution from water Cerenkov track
reconstruction depends on the photo-multiplier tube coverage.  With
coverage greater than $\sim$ 10\%, we expect that the
reconstruction resolution should be adequate for our purposes. In the
following discussion 
we assume a 10\% resolution on the $\nu_\mu$ energy.  This
is consistent with the resolution achieved by the K2K experiment.

The range of $\mdmatm \sim 1.24{E_\nu\mbox{[\GeV{}]} \over
  L\mbox{[km]}}$ covered by the proposed experiment using the beam in
Fig. \ref{bnlspec} extends to the low value of about $5 \times 10^{-4}
\ \meV^2$.  The lower end of this extensive range of values is
considerably below the corresponding values for other  long
baseline terrestrial experiments~\cite{minos,cngs}.  If the value of
\dmatm{} turns out to be towards the lower end ($\sim 10^{-3}$) of its
current range or if the value of \dmsol{} turns out to be towards its
high end ($\sim$ $10^{-4} \meV^2$) then  large and very
interesting interference effects in the very long baseline experiment
will be possible. 

\begin{figure}
  \begin{center}
    \includegraphics*[width=\textwidth]{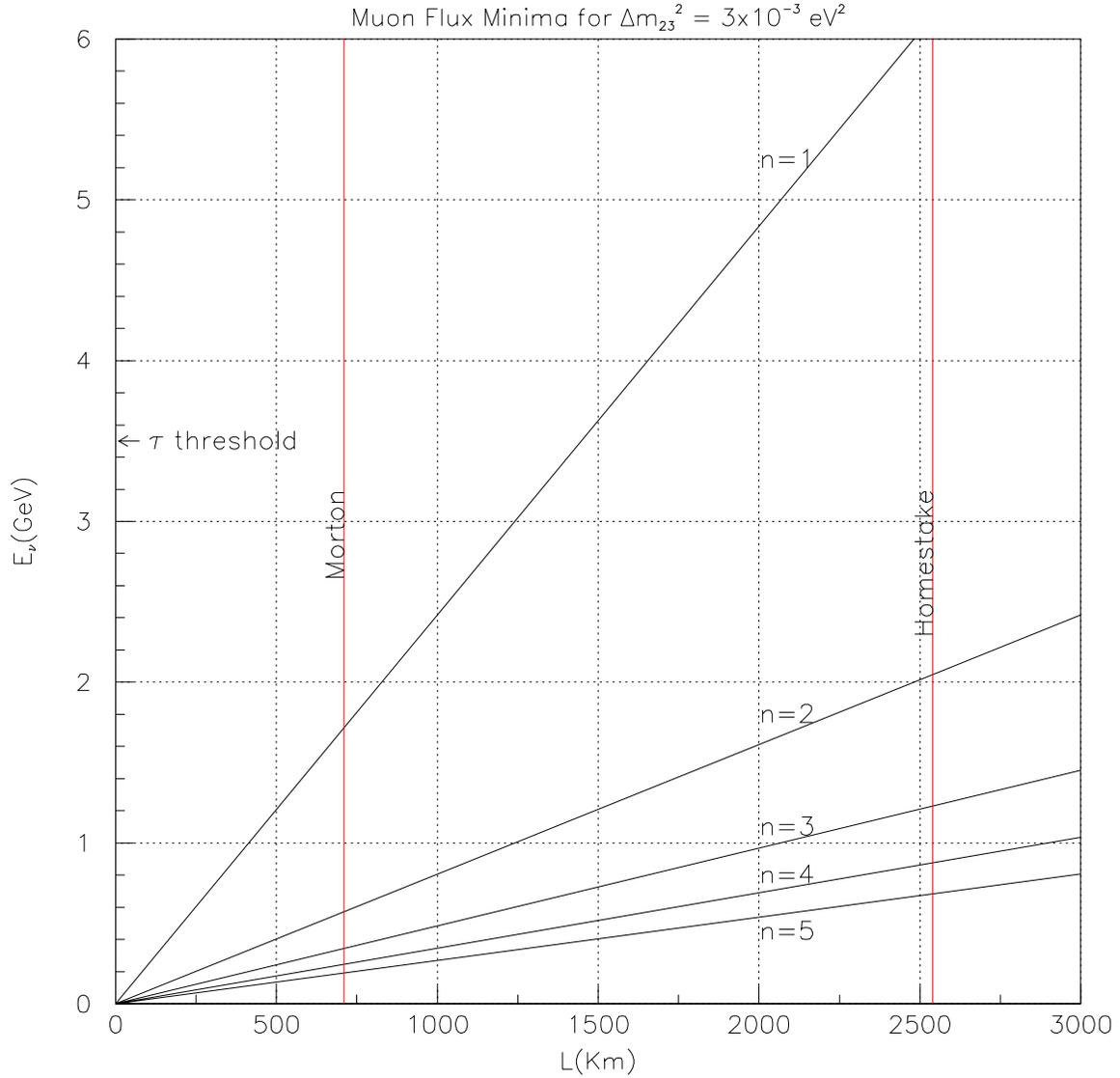} 
    \caption{Nodes of neutrino oscillations as a function of oscillation length and
      energy for $\mdmatm = 0.003 \meV^2$.  Matter effects are not 
included. The distance to Morton salt works (location of the old 
IMB experiment \cite{imb}) and Homestake is shown by the vertical lines. 
}
    \label{nodes}  
  \end{center}
\end{figure}

\begin{figure}
  \begin{center}
    \includegraphics*[width=\textwidth]{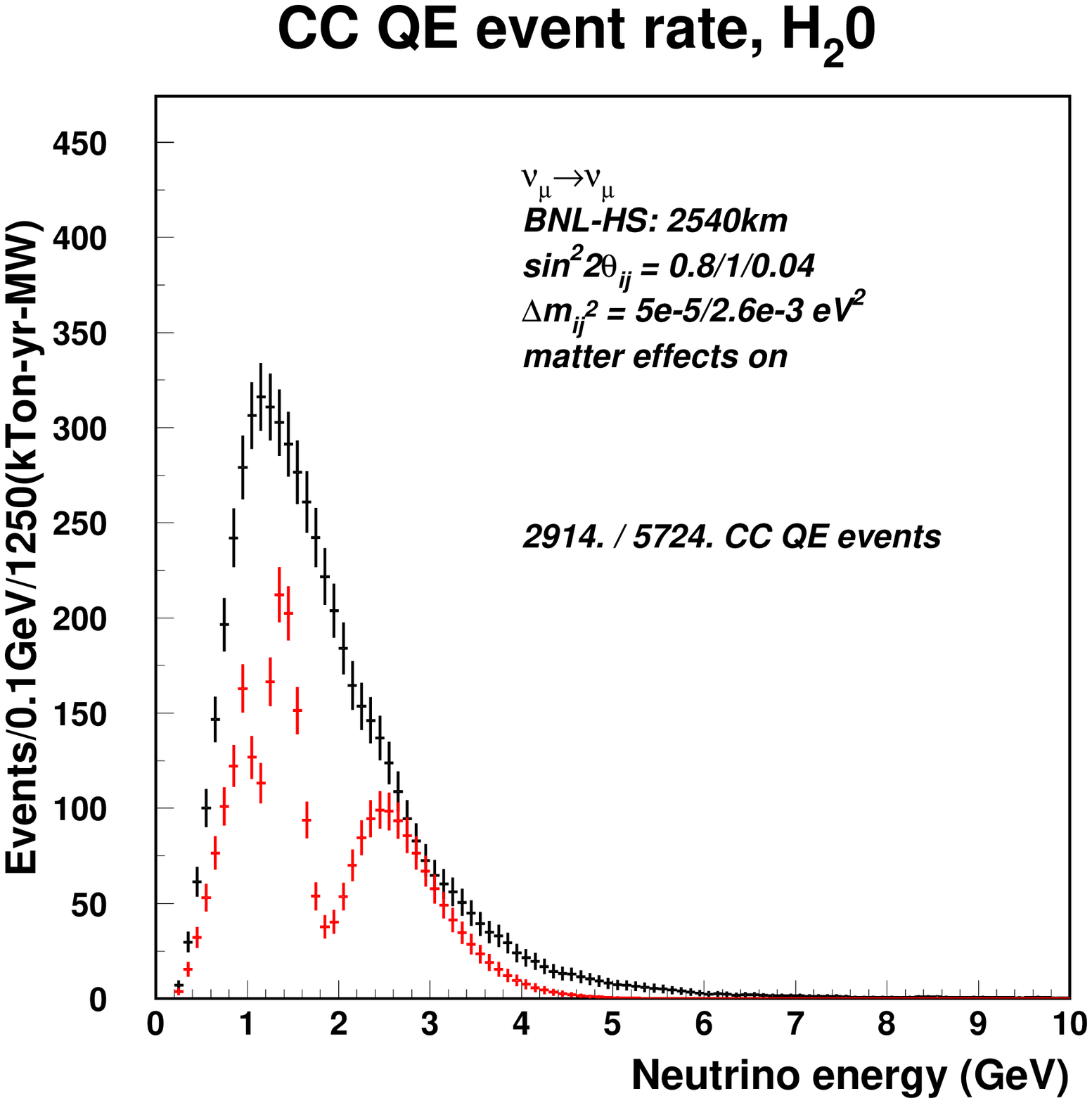}
    \caption{Spectrum of detected quasi-elastic events in a 0.5 MT detector at
      2540 km from BNL. We have assumed 0.5 MW of beam power and 5
      years of running.  The top data points are without oscillations
      and bottom are with oscillations.  This plot is for $\mdmatm =
      0.0026 \meV^2$.
      The error bars correspond to the statistical error expected in
      the bin. A 10 \% energy resolution is assumed; this corresponds
      to the expected resolution due to both nuclear effects and the
      muon momentum reconstruction in the detector.}
    \label{wcnodesa}  
  \end{center}
\end{figure}

\begin{figure}
  \begin{center}
    \includegraphics*[width=\textwidth]{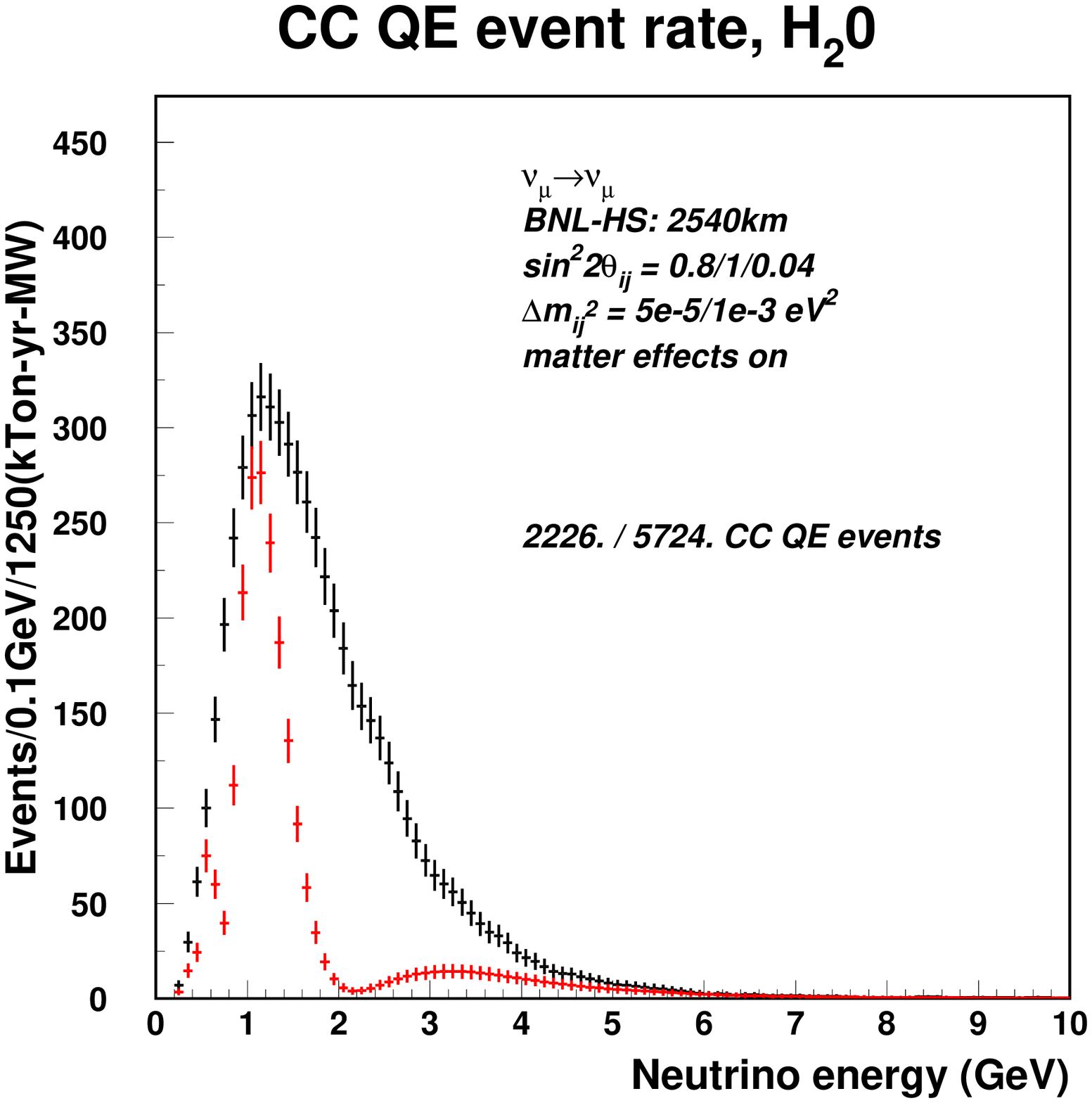}
    \caption{Spectrum of detected quasi-elastic events in a 0.5 MT detector at
      2540 km from BNL. We have assumed 0.5 MW of beam power and 5
      years of running.  The top data points are without oscillations
      and bottom are with oscillations. This plot is for $\mdmatm = 0.001
      \meV^2$.  The error bars correspond to the statistical error
      expected in the bin. A 10 \% energy resolution is assumed; this
      corresponds to the expected resolution due to both nuclear
      effects and the muon momentum reconstruction in the detector.}
    \label{wcnodesb}  
  \end{center}
\end{figure}

Extra-long neutrino flight paths open the possibility of observing
multiple nodes (minimum intensity points) of the neutrino oscillation
probability in the disappearance experiment.  Observation of one such
pattern will for the first time directly demonstrate the oscillatory
nature of the flavor changing phenomenon.  The nodes occur at
distances $L_n = 1.24 (2n-1) E_{\nu}/\mdmatm$, $n= 1,2,3, $ \ldots.
In Fig. \ref{nodes}, as an example, we show the flight path $L$ versus
$E_{\nu}$ relationship of the nodes for $\Delta m^2 = 0.003 {\rm
  e\mbox{V}}^2$, a value close to the value measured in atmospheric
neutrino experiments \cite{sk}.  An advantage of having a very
long baseline is that the multiple node pattern is detectable over a
broad range of $\Delta m^2$.  This is demonstrated in figures
\ref{wcnodesa} and \ref{wcnodesb} which shows the disappearance of
muon type neutrino events as a function of neutrino energy 
 in quasi-elastic events; the plots do
not contain background.  For \dmatm{} as small as 0.001 $e\mbox{V}^2$
the oscillation effects will be very large.  Even in the presence of
background we expect that the statistical error on \dmatm{} will be
small ($\sim$ 0.5 \%) and the systematic error from the energy scale
of the detector will dominate the total error.


The oscillation of \numunue{} is 
discussed is several recent papers \cite{arafune, marciano, irina}. 
This oscillation in vacuum is  described fully by 
the following equation:  
\begin{eqnarray}
  \label{eq:one}
P(\nu_\mu\to\nu_e) & = & 4(s^2_2s^2_3c^2_3 +J_{CP}\sin\Delta_{21})
\sin^2\frac{\Delta_{31}}{2} \nonumber \\
& & +2(s_1s_2s_3c_1c_2c^2_3 \cos\delta -s^2_1s^2_2s^2_3c^2_3) \sin
\Delta_{31} \sin \Delta_{21} \label{eq9} \\
& & +4(s^2_1c^2_1c^2_2c^2_3 +s^4_1s^2_2s^2_3c^2_3 -2s^3_1s_2s_3c_1c_2c^2_3
\cos\delta -J_{CP} \sin\Delta_{31}) \sin^2\frac{\Delta_{21}}{2}
\nonumber \\
& & +8(s_1s_2s_3c_1c_2c^2_3 \cos\delta - s^2_1s^2_2s^2_3c^2_3) \sin^2
\frac{\Delta_{31}}{2} \sin^2 \frac{\Delta_{21}}{2} \nonumber
\end{eqnarray}
where 
\begin{equation}
  \label{eq:blah}
  J_{CP} \equiv s_1s_2s_3c_1c_2c^2_3\sin\delta \label{eq6}
\end{equation}

\begin{figure}
  \begin{center}
    \includegraphics*[width=\textwidth]{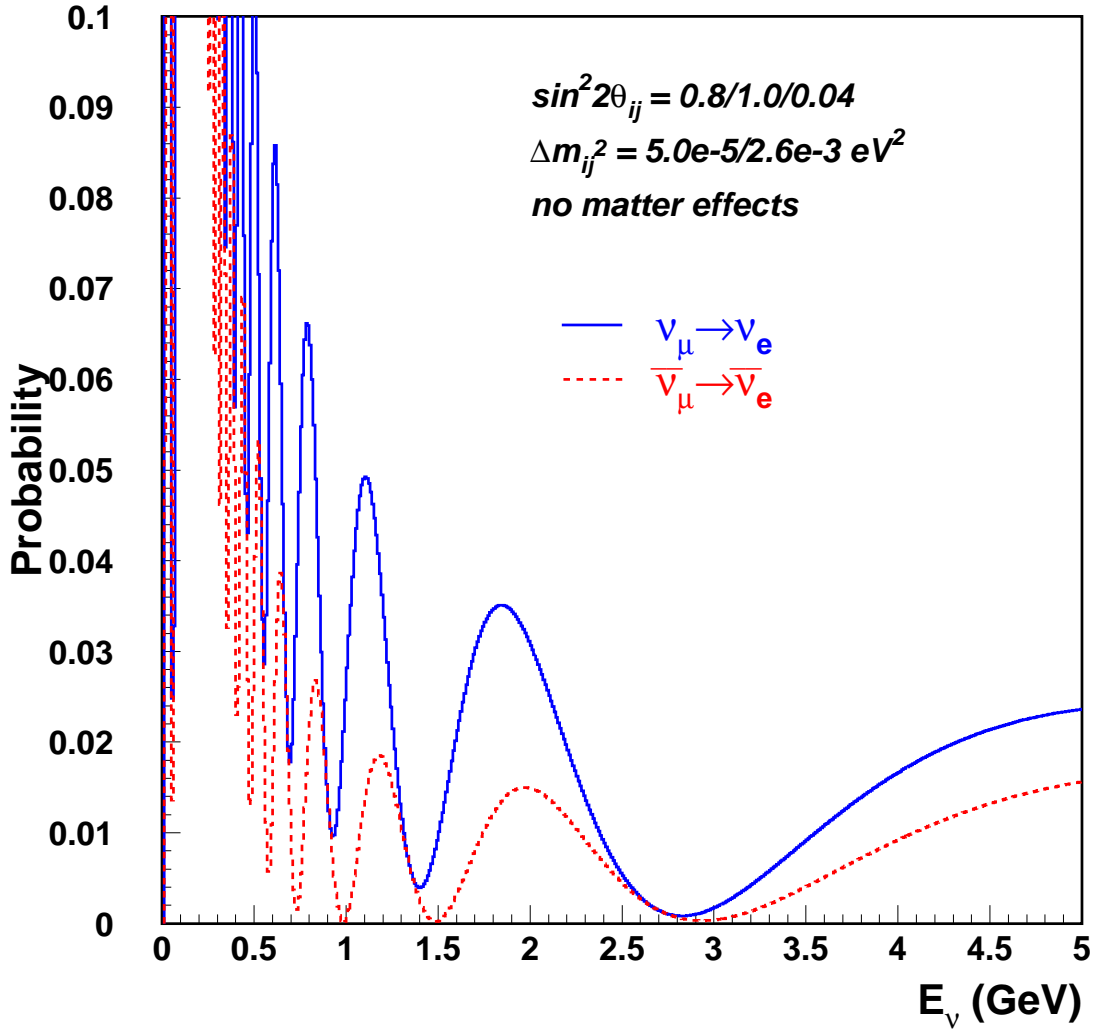} \\
    \caption{Probability of \numunue{}
      and \anumunue{}  oscillations at 2540 km
      assuming a $45^o$ CP violation phase. It can be seen that the 
      CP asymmetry between $\nu_\mu$ and $\bar\nu_\mu$ increases
      for lower energies because the CP asymmetry is proportional
      to $\mdmsol L /E$ which increases for lower energies. 
    }
    \label{cpasym}
  \end{center}
\end{figure}

$J_{CP}$ is an invariant that quantifies CP violation in the neutrino
sector. The formula for $P(\bar\nu_\mu\to\bar\nu_e)$ is the same as
above except that the $J_{CP}$ terms have opposite sign.  Please see
attached appendix (hep-ph/0108181) for definitions of symbols but note
that $\Delta_{31}$ is the atmospheric term and $\Delta_{21}$ is the
solar term. The vacuum oscillations for a baseline of 2540 km are
illustrated in Fig.~\ref{cpasym} as a function of energy for both muon
and anti-muon neutrinos.  The main feature of the oscillation is due
to the term linear in $\sin^2\frac{\Delta_{31}}{2}$. The oscillation
probability rises for lower energies due to the terms linear in
$\sin^2 \frac{\Delta_{21}}{2}$.  The interference terms involve CP
violation and they create an asymmetry between neutrinos and
anti-neutrinos.  The vacuum oscillation formula (Eq.\ref{eq:one})
  and Fig.~\ref{cpasym}
show that the CP asymmetry also grows as $1/E$ in the 0.5-3.0 \GeV{}
region.  Because of this effect it is argued that the figure of merit
for measuring CP violation is independent of the baseline. For very
long baselines  the statistics for a given size detector at a
given energy are poorer by one over the square of the distance, but 
the CP asymmetry grows linearly in distance \cite{marciano}.

\begin{figure}
  \begin{center}
    \includegraphics*[width=\textwidth]{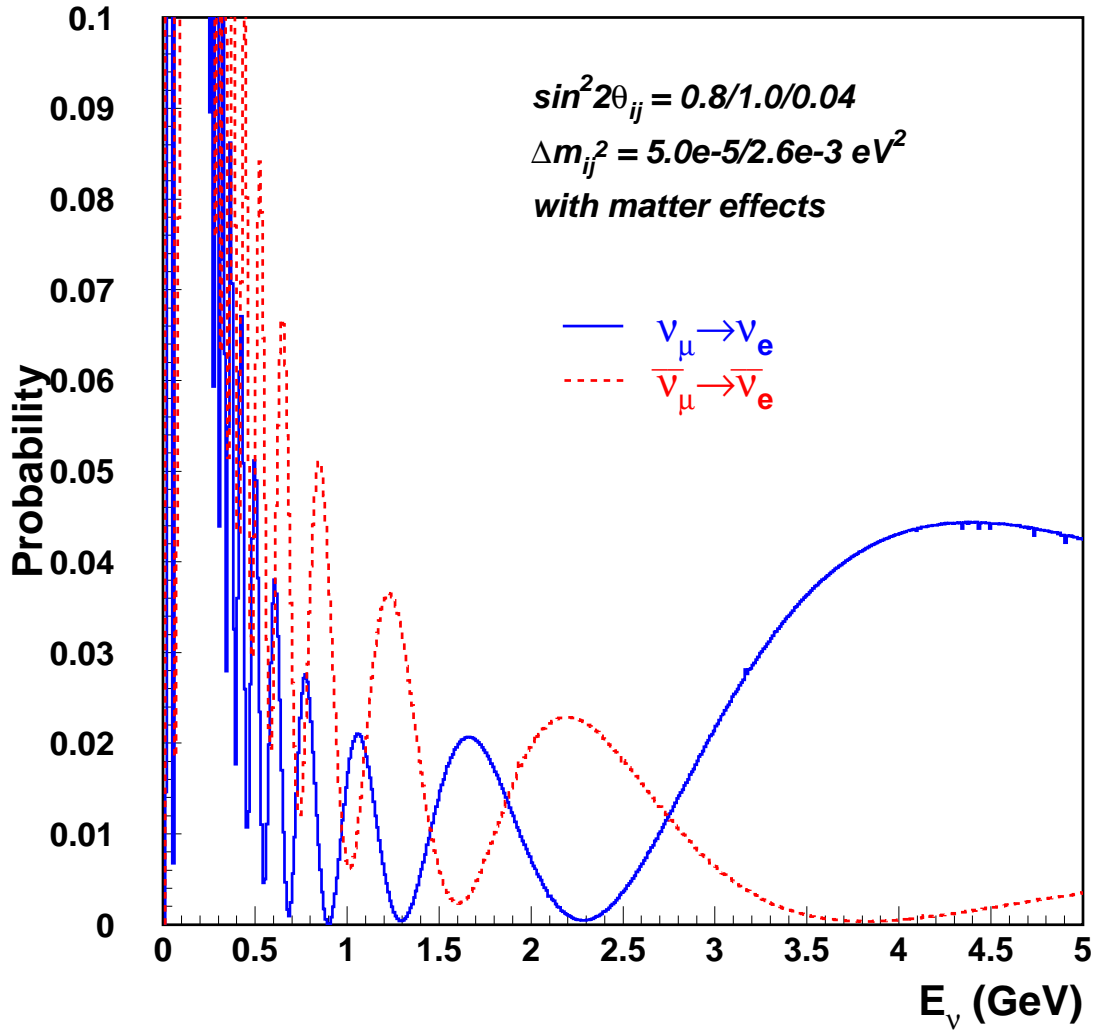}
    \caption{Probability of $\nu_\mu$ oscillating into 
      $\nu_e$ after 2540 km.  The parameters assumed are listed in the
      figures.  This plot assumes that there is no CP violation in
      the neutrino mixing matrix.}
    \label{pnumunuea}
  \end{center}
\end{figure}

\begin{figure}
  \begin{center}
    \includegraphics*[width=\textwidth]{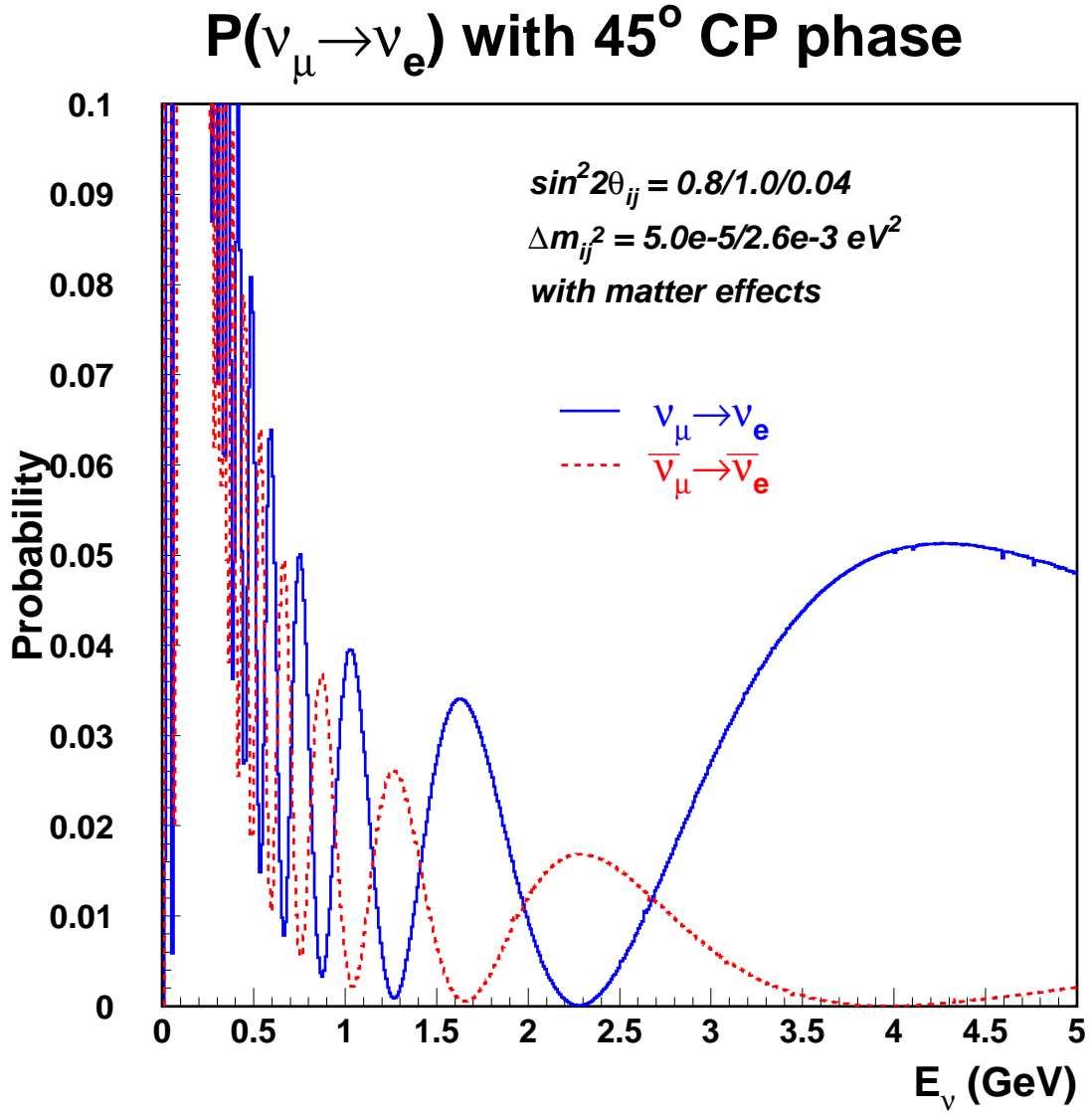}
    \caption{Probability of $\nu_\mu$ oscillating into 
      $\nu_e$ after 2540 km.  The parameters assumed are listed in the
      figures.  This plot assumes a CP violation phase of $45^o$.  }
    \label{pnumunueb}
  \end{center}
\end{figure}

The vacuum oscillation formulation must be modified to include the
effect of matter \cite{irina}.  The \numunue{} probability in the
presence of matter is shown in Figs.~\ref{pnumunuea} and
\ref{pnumunueb}. When compared to Fig.~\ref{cpasym} we can see that
matter will enhance (suppress) neutrino (anti-neutrino) conversion at
high energies and will also lower (increase) the energy at which the
oscillation maximum occurs.  The effect is opposite (enhancement for
anti-neutrinos and suppression for neutrinos) if the sign of \dmatm{}
is negative.  The matter enhancement effect in neutrino oscillations
has been postulated for a long time without experimental
confirmation~\cite{wolfenstein}.  Detection of such an effect by
measuring a large asymmetry between neutrino and anti-neutrino
oscillations or by measuring the spectrum of electron neutrinos is a
major goal for neutrino physics. This measurement will also yield the
sign of \dmatm{}.

\begin{figure}
  \begin{center}
    \includegraphics*[width=\textwidth]{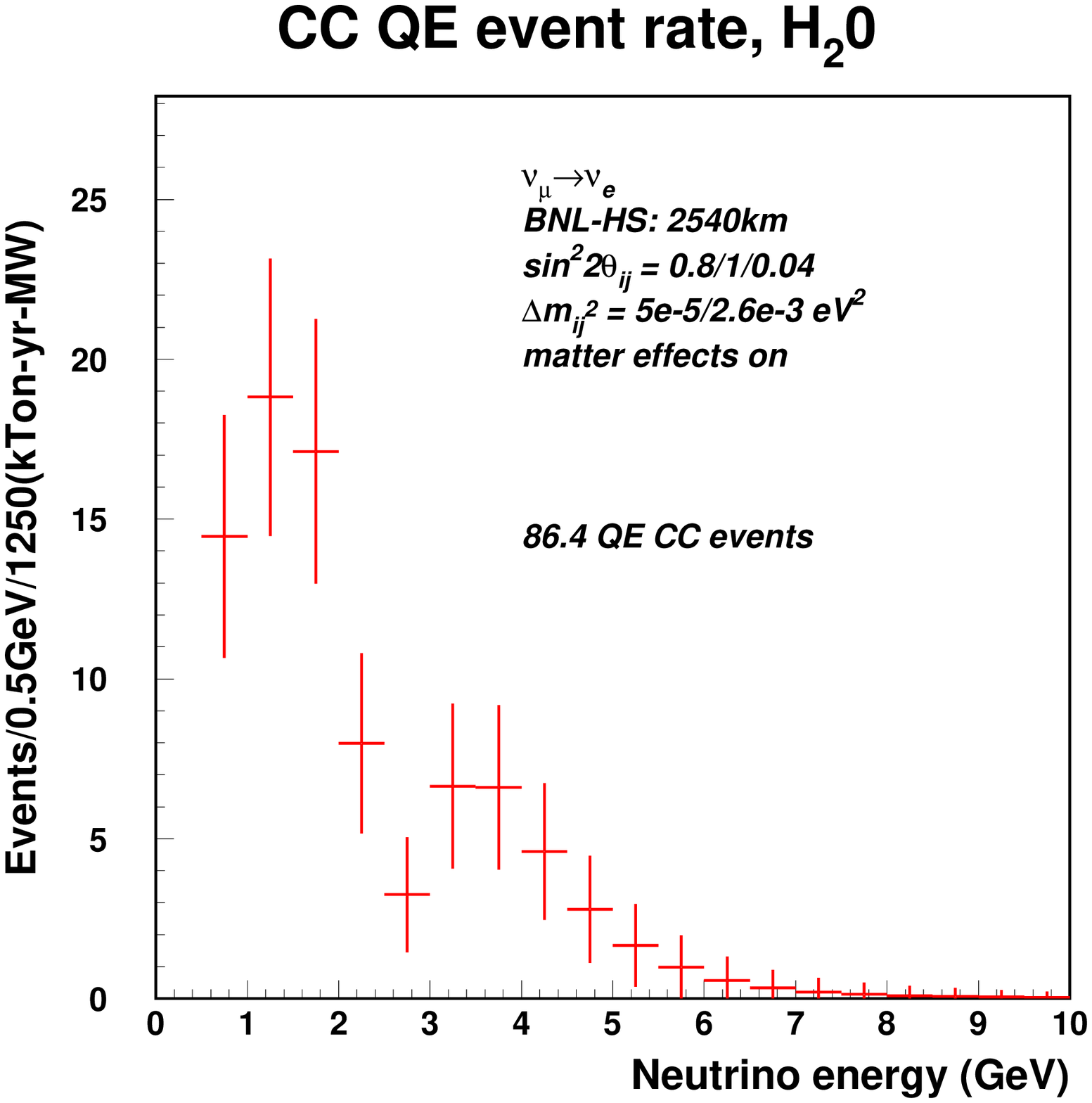}
    \caption{Spectrum of detected quasi-elastic electron neutrino 
      charged current events in a 0.5 MT detector at 2540 km from BNL.
      We have assumed 0.5 MW of beam power and 5 years of running.
      This plot is for $\mdmatm = 0.0026 \meV^2$.  We have assumed
      $\sin^2 2 \theta_{13} = 0.04$ and 
$\mdmsol = 5\times 10^{-5} \meV^2$. 
 The error bars correspond to the
      statistical error expected in the bin. A 10 \% energy resolution
      is assumed; this corresponds to the expected resolution due to
      both nuclear effects and the electron momentum reconstruction in
      the detector.}
    \label{nuenodesa}
  \end{center}
\end{figure}

\begin{figure}
  \begin{center}
    \includegraphics*[width=\textwidth]{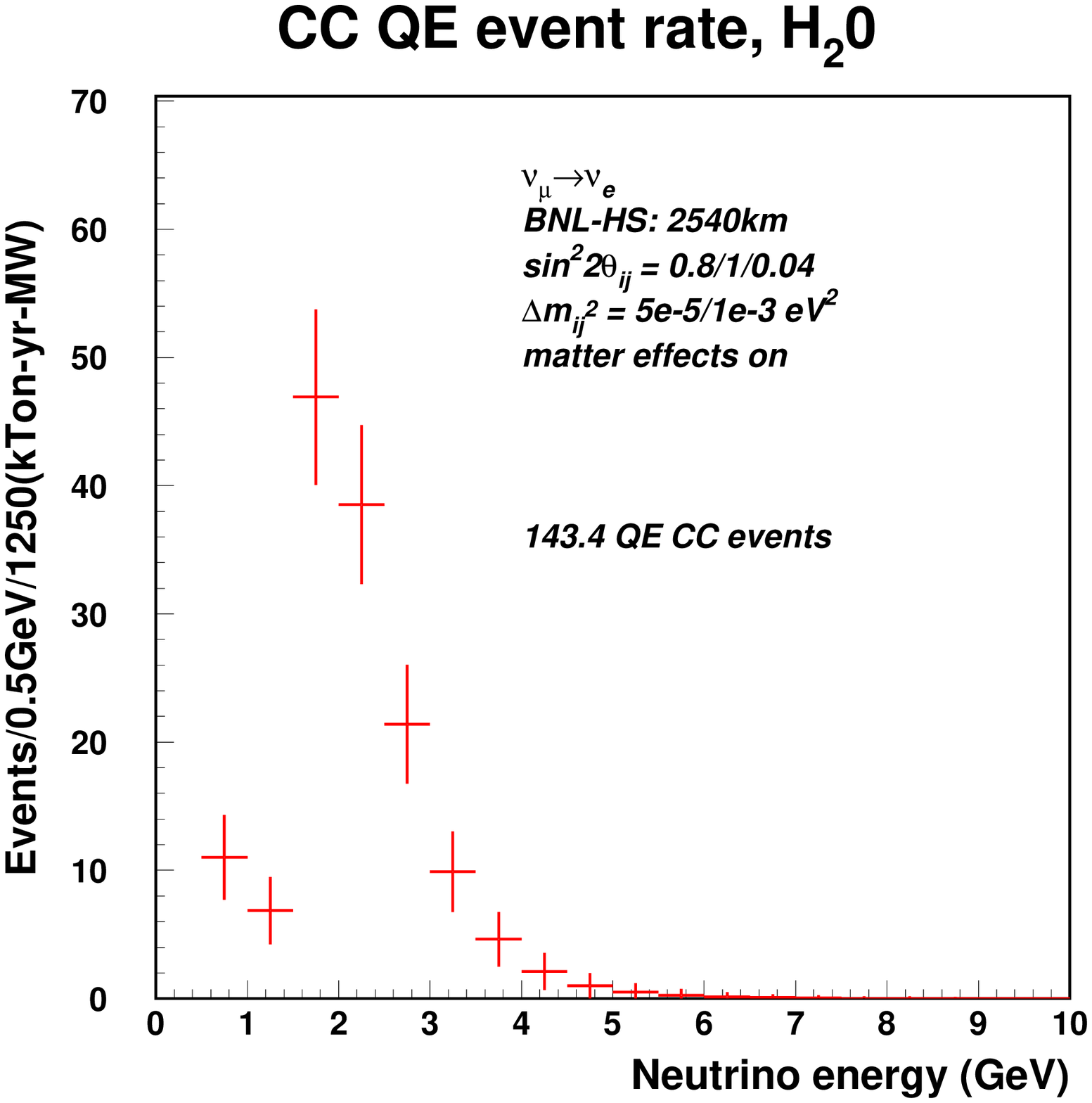}
    \caption{Spectrum of detected quasi-elastic electron neutrino 
      charged current events in a 0.5 MT detector at 2540 km from BNL.
      We have assumed 0.5 MW of beam power and 5 years of running.
      This plot is for $\mdmatm = 0.001 \meV^2$.  We have assumed
      $\sin^2 2 \theta_{13} = 0.04$ and  
$\mdmsol = 5\times 10^{-5} \meV^2$. 
The error bars correspond to the
      statistical error expected in the bin. A 10 \% energy resolution
      is assumed; this corresponds to the expected resolution due to
      both nuclear effects and the electron momentum reconstruction in
      the detector.}
    \label{nuenodesb}
  \end{center}
\end{figure}

\begin{figure}
  \begin{center}
    \includegraphics*[width=\textwidth]{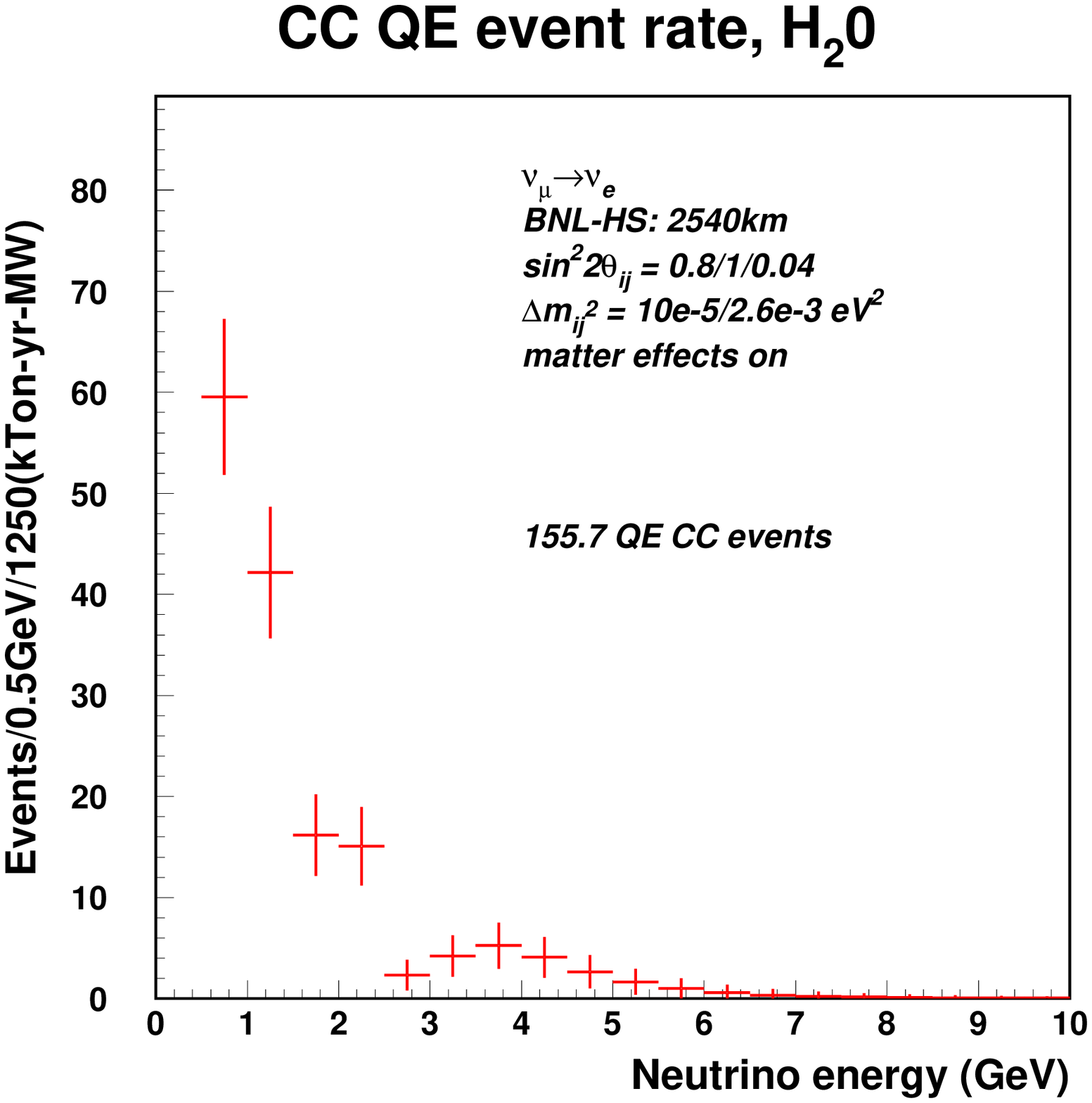}
    \caption{
      Spectrum of detected quasi-elastic electron neutrino 
      events in a 0.5 MT detector at
      2540 km from BNL. We have 
assumed 0.5 MW of beam power and 5 years of running. 
      This plot is for $\mdmatm = 0.0026 \meV^2$ and 
      $\mdmsol = 0.0001 \meV^2$.
      We have assumed  $\sin^2 2 \theta_{13} = 0.04$. 
      The error bars correspond to the statistical error expected in the bin. A 10 \% 
      energy resolution is assumed; this corresponds to the expected resolution due to 
      both nuclear effects and the electron momentum reconstruction. 
    }
    \label{lm12}
  \end{center}
\end{figure}

Figures \ref{nuenodesa}, \ref{nuenodesb} and \ref{lm12} show the spectrum of electron
type neutrinos that will be detected at 2540 km.  The signal for
$\sin^2 2 \theta_{13} \sim 0.04$ will be about 100 events. The
background for this signal will come from the intrinsic contamination
of $\nu_e$ particles in the beam as well as neutral current events
producing $\pi^0$s.  This background will be examined in detail in a
future update to this proposal. From past experience using this beam,
we expect that the total background in the signal region can be
reduced to about 0.5\% of the charged current muon neutrino events.
The advantages of the very long baseline are in obtaining a large
enhancement at higher energies and creating a nodal pattern in the
appearance spectrum. Both of these can be used to further improve the
sensitivity of the experiment. The very long baseline experiment has a
great advantage if \dmsol{} is found to be somewhat larger within its
allowed range $(2-10) \times 10^{-5} \meV^2$. This is shown in
Fig.~\ref{lm12}.  The differences in the electron neutrino spectra are
striking within Figs.~\ref{nuenodesa} and \ref{nuenodesb},\ref{lm12}.

To understand CP violation, the effect of matter enhancement must be
clearly understood and subtracted from any observation.  If CP
violation is large and the signal to \numunue{} is also large then it
is possible to measure CP violation with just the ($\nu_\mu$) 
neutrino beam. As
shown in Fig.~\ref{cpasym} the effects of CP violation grow linearly
as energy is decreased (or the baseline increased). For a very long
baseline experiment it is possible to compare the signal strength in
the $\pi/2$ node versus the $3\pi/2$ or higher nodes. Such a
comparison will yield a measurement of CP violation.  Any such
measurement of CP must be augmented by data using a muon
 anti-neutrino
beam.  Such a program of measurements will require large statistics.
This proposal has the flexibility to obtain much larger data sets
because the detector will eventually be upgraded to its final
configuration with 1 MT of mass and the AGS accelerator complex can be
upgraded up to 2.5 MW of beam power.  It is also possible that the
conventional neutrino beam which we propose here will be replaced by
a neutrino factory based on a muon storage ring \cite{study2}.




\subsection{Detectors for the very long baseline experiment} 

The conversion of Homestake Gold Mine in Lead, South Dakota, into the
National Underground Science Laboratory (NUSL), tentatively to take
place in 2002, will provide a unique opportunity for a program of
extra-long baseline neutrino oscillation experiments.  As explained
above these will be possible because of the length of the baseline,
2540 km from the Brookhaven National Laboratory (BNL) to Lead, South
Dakota.  It is proposed that the NUSL facility will accommodate an
array of detectors with total mass approaching 1 Megaton. Most of
these will be water Cerenkov detectors that can observe neutrino
interactions in the desired energy range with sufficient energy and
time resolution \cite{3m}.

An alternative to Homestake also exists at the Waste Isolation Pilot
Plant (WIPP) located in an ancient salt bed at a depth of $\sim 700 m$
near Carlsbad, New Mexico. The distance from BNL to WIPP is about 2880
km.  The cosmic ray background will be higher at WIPP because the
facility is not as deep as Homestake which has levels as deep as $\sim
2500 m$.  The increased background, although undesirable, is not an
insurmountable problem. However, the mechanical design of a large
cavity in a salt bed has to be very different because of the slow
movement of salt that causes a cavity to slowly collapse in a salt
mine. In this LOI we will not address the detailed issues of detector
design and cost.  A more detailed study of a very large water Cerenkov
detector has been done by the UNO collaboration \cite{uno}.

\begin{figure}
  \begin{center}
    \includegraphics*[width=\textwidth]{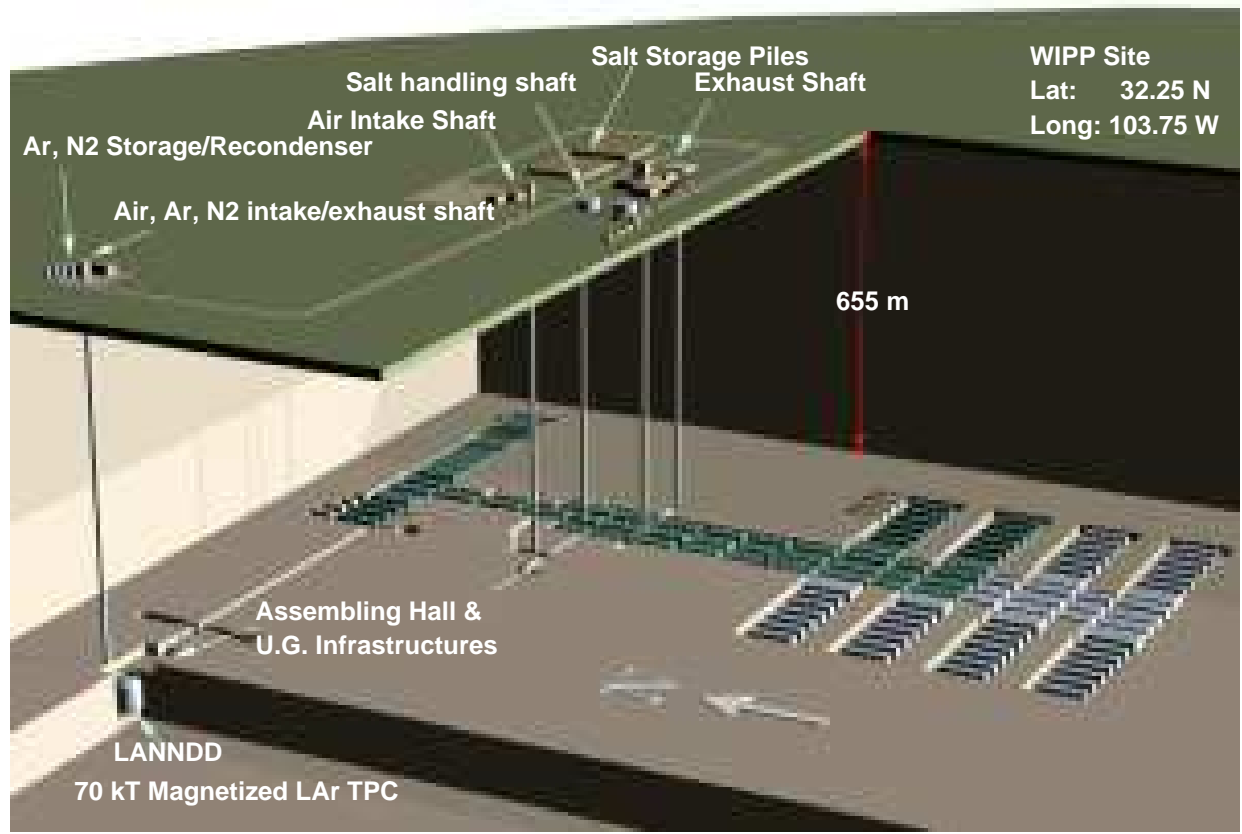} 
    \caption{Schematic of the WIPP underground site and a location 
      for the LANNDD detector.  A water Cerenkov detector array could
      also be accommodated at the WIPP site. }
    \label{landd}
  \end{center}
\end{figure}

Currently, a study is in progress to site the LANNDD, 70 KT liquid
Argon detector at WIPP \cite{lannddp}. The key issue at this stage is one of safety and
a proposal to the DOE to study this is in preparation.  In Figure
\ref{landd}, we show the LANNDD detector in a possible underground
location at WIPP.  One advantage of the WIPP site is that, it is owned
by the DOE and now has a program of underground science.
We note that the recent Neutrino Factory Study \cite{study2}
 at BNL identified the
WIPP site as one possible location for a far detector,
and the current BNL neutrino beam could use the same concept.  The
LANNDD detector can be used for  neutrino physics, as well as 
the search for proton decay and other astro-particle
physics goals.  Currently, the ICARUS detector at the Gran Sasso is
being constructed with a 3KT detector as a goal.  The operation of
this detector will provide key information for the eventual
construction of LANNDD and for the neutrino physics identified in this
LOI.

\vspace{1ex}

\section{Long Baseline Experiment}

\begin{figure}
  \begin{center}
    \includegraphics*[width=\textwidth]{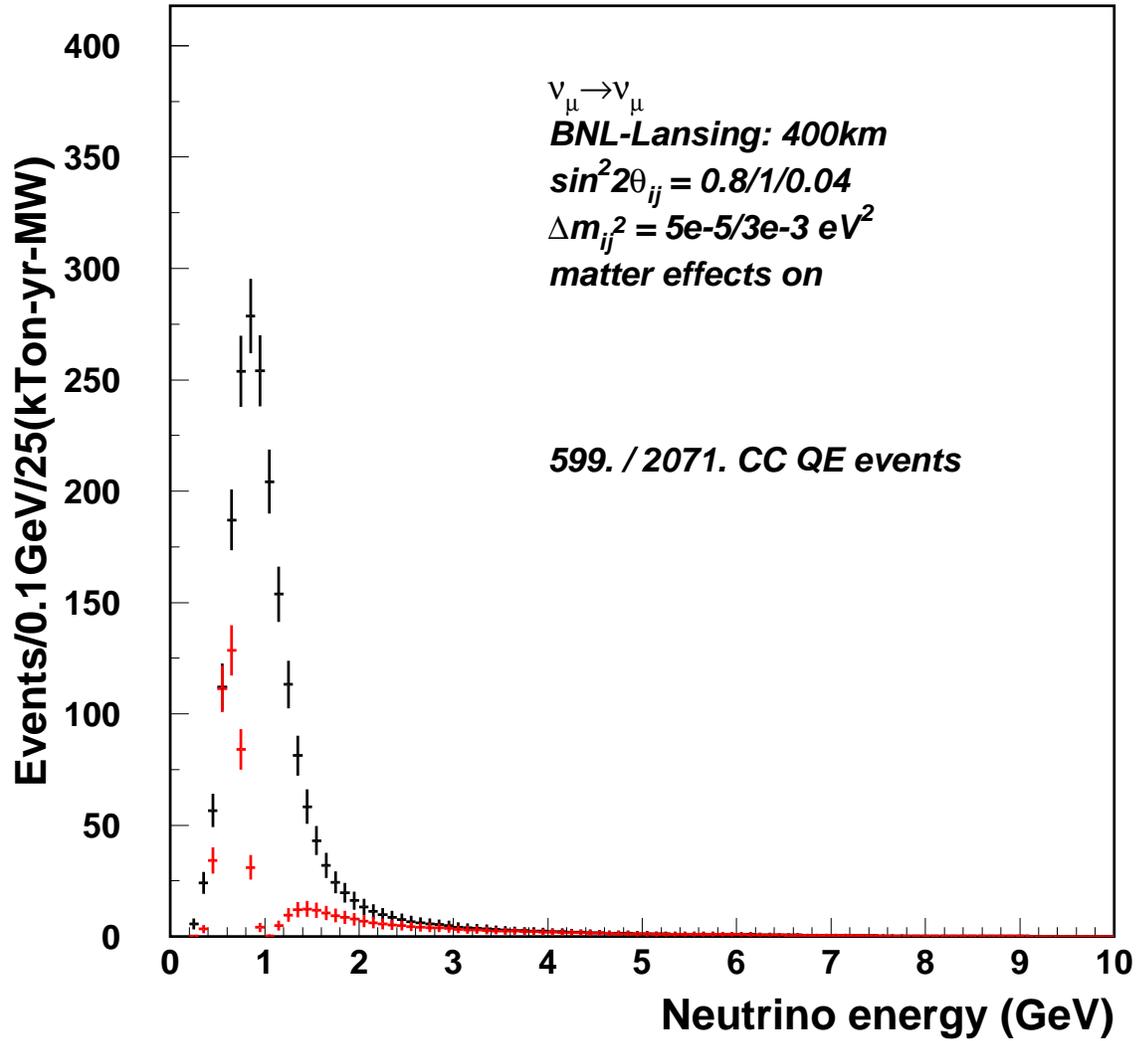}
    \caption{Spectrum of neutrinos at 1.5 degrees with and without oscillations with 
      $\mdmatm = 0.003 \meV^2$ and full mixing.}
    \label{400km}
  \end{center}
\end{figure}

A dedicated experiment to detect \numunue{} appearance signature in
the \dmatm{} region of 0.003 $\meV^2$ needs to have low background
from the intrinsic $\nu_e$ contamination as well as neutral current
events that produce $\pi^0$s which can mimic the $\nu_e$ QE CC
signature.  This can be achieved by placing a fine grained detector
1.5 degrees off-axis from the BNL neutrino beam.  As originally
pointed out by the E889 collaboration \cite{e889} at angles larger
than the divergence of the pion beam the neutrino spectrum is almost
independent of the pion energy and has a narrow spectrum peaking at 1
\GeV{} (See Fig.~\ref{bnlspec}).  If the value of $\mdmatm = 0.003
\meV^2$ is known with good precision then the detector could be placed
at 412 km which is at the point of the first maximum for oscillations.
This is shown in figure \ref{400km}.  At this distance and energy the
effect of matter enhancement is small and a very sensitive experiment
for \numunue{} appearance could be performed with a high resolution
fine grained detector such as a  liquid argon time projection 
chamber, with maximum performance achieved if the detector is
immersed in a $\sim$ 1/2 T magnetic field \cite{lannddp}. 

 We calculate the
number of events in a 1.5 degree off-axis beam in a 10 kT liquid Argon
detector assuming 0.5 MW of beam power and 5 years of running. In the
absence of oscillations there will be about 2000 quasi-elastic muon
neutrino events. The total number of events including neutral current
events will be about 1.5 times this number. If $\mdmatm = 0.003
\meV^2$ the number of muon quasi-elastic events will drop to only
$\sim$200.  If $\sin^2\theta_{13} = 0.01$ then the number of electron
neutrino quasi-elastic events will be about 40. The background from the
electron neutrino contamination in the beam and the neutral current
$\pi^0$ will be spread widely in energy, but the signal of 40 events
will be in the peak of the neutrino spectrum.  
The high energy resolution of the liquid Argon
detector as well as the granularity should make it possible to reduce
the neutral current background to a negligible level \cite{larhighres}. 
 The intrinsic
electron neutrino background in the AGS neutrino beam is known to be
at 0.5 \% level \cite{e889}.

\begin{figure}[htp]  
  \begin{center}
    \includegraphics*[width=\textwidth]{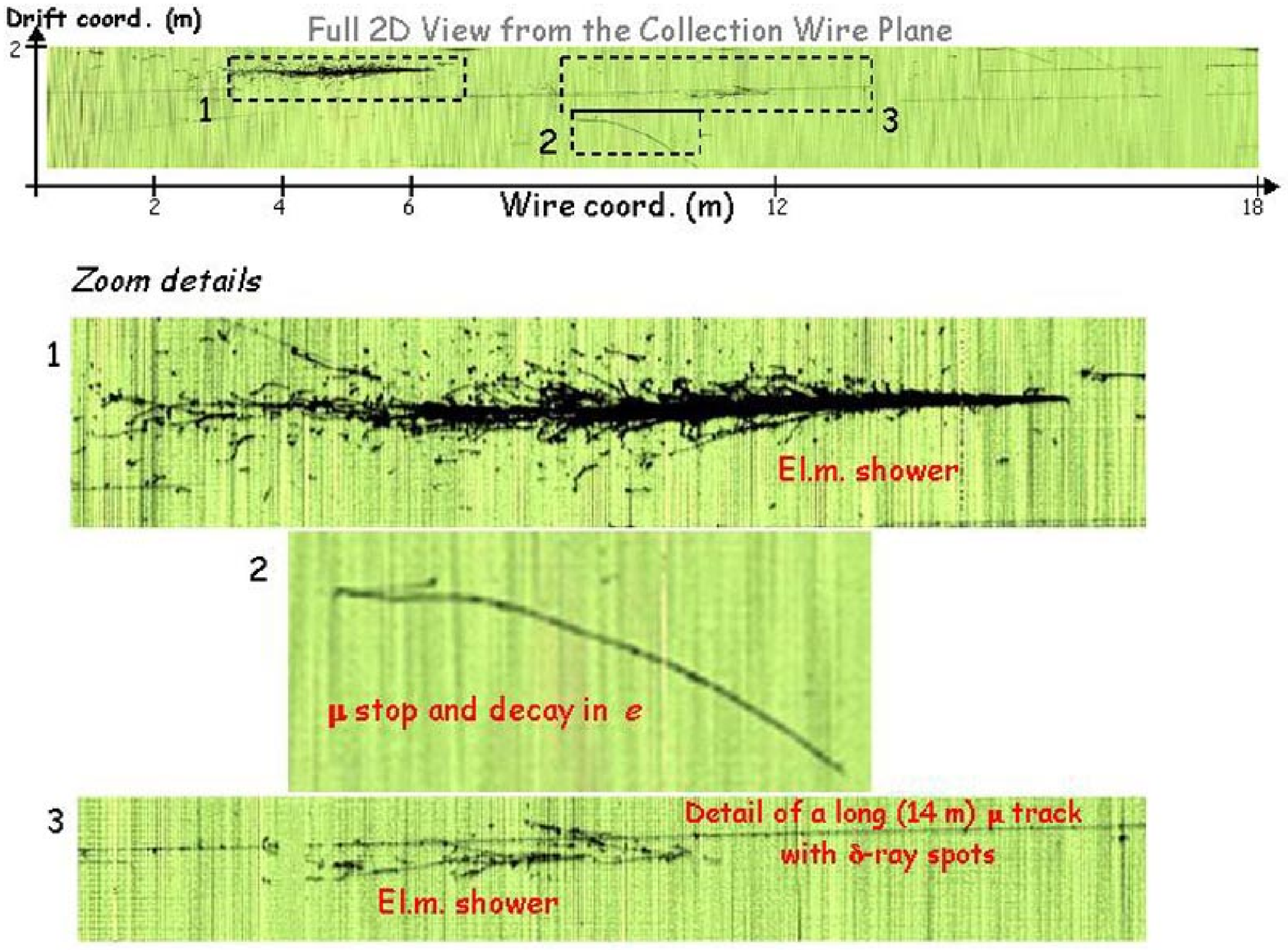}
    \parbox{5.5in} 
    {\caption[ Short caption for table of contents ]
      {\label{icarus2} An event from the recent cosmic-ray test run of ICARUS  
        \cite{ICARUS},
        showing excellent track resolution over long drift distances in zero magnetic 
        field.
      }}
  \end{center}
\end{figure}

The viability of a large liquid argon detector is presently being demonstrated
by the ICARUS collaboration \cite{ICARUS} in cosmic-ray tests of a 
300-ton module located on the Earth's surface.  Fig.~\ref{icarus2} shows
an example of the detailed tracking information obtainable with this
technology.

A magnetized liquid argon detector would give the maximal discrimination
against backgrounds in a neutrino beam, would enhance the ability to
perform CP violation experiments, and would permit use of a beam
produced by a solenoid focussing scheme \cite{skahn}
that contains both neutrinos and antineutrinos.  An R\&D experiment is
proposed to use a prototype liquid argon detector in
a magnetic field to determine the sign of electrons via analysis of
their electromagnetic showers up to several \GeV{} \cite{argonprop}.

\section{AGS Upgrade}

\vspace{1ex}

\begin{figure}
  \begin{center}
    \includegraphics*[width=\textwidth]{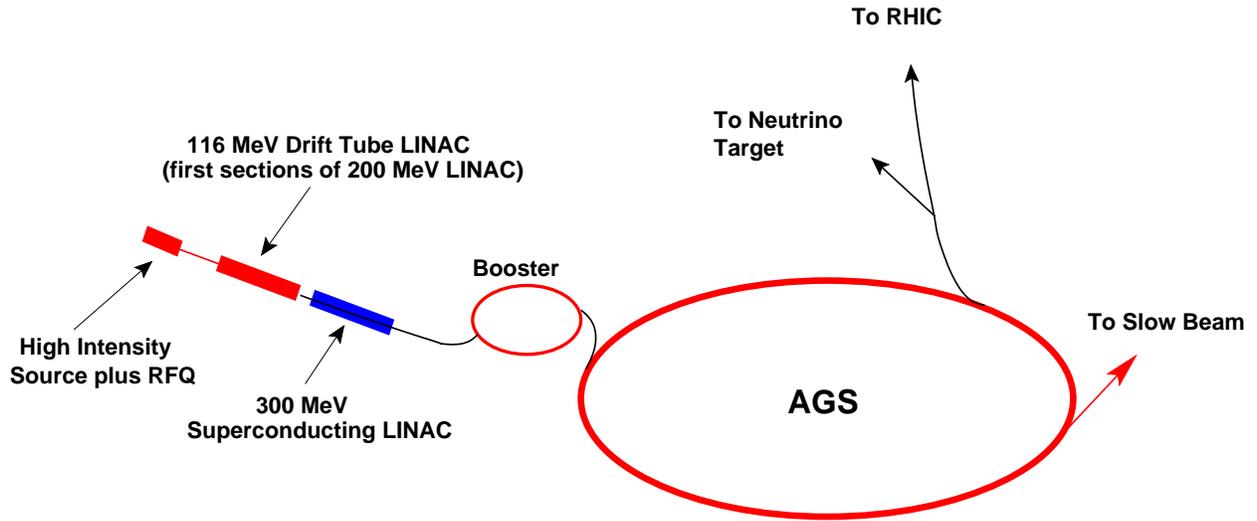}
    \caption{Layout of the AGS facility with the addition of the super conducting LINAC.} 
    \label{page3}
  \end{center}
\end{figure}

\begin{figure}
  \begin{center}
    \includegraphics*[width=\textwidth]{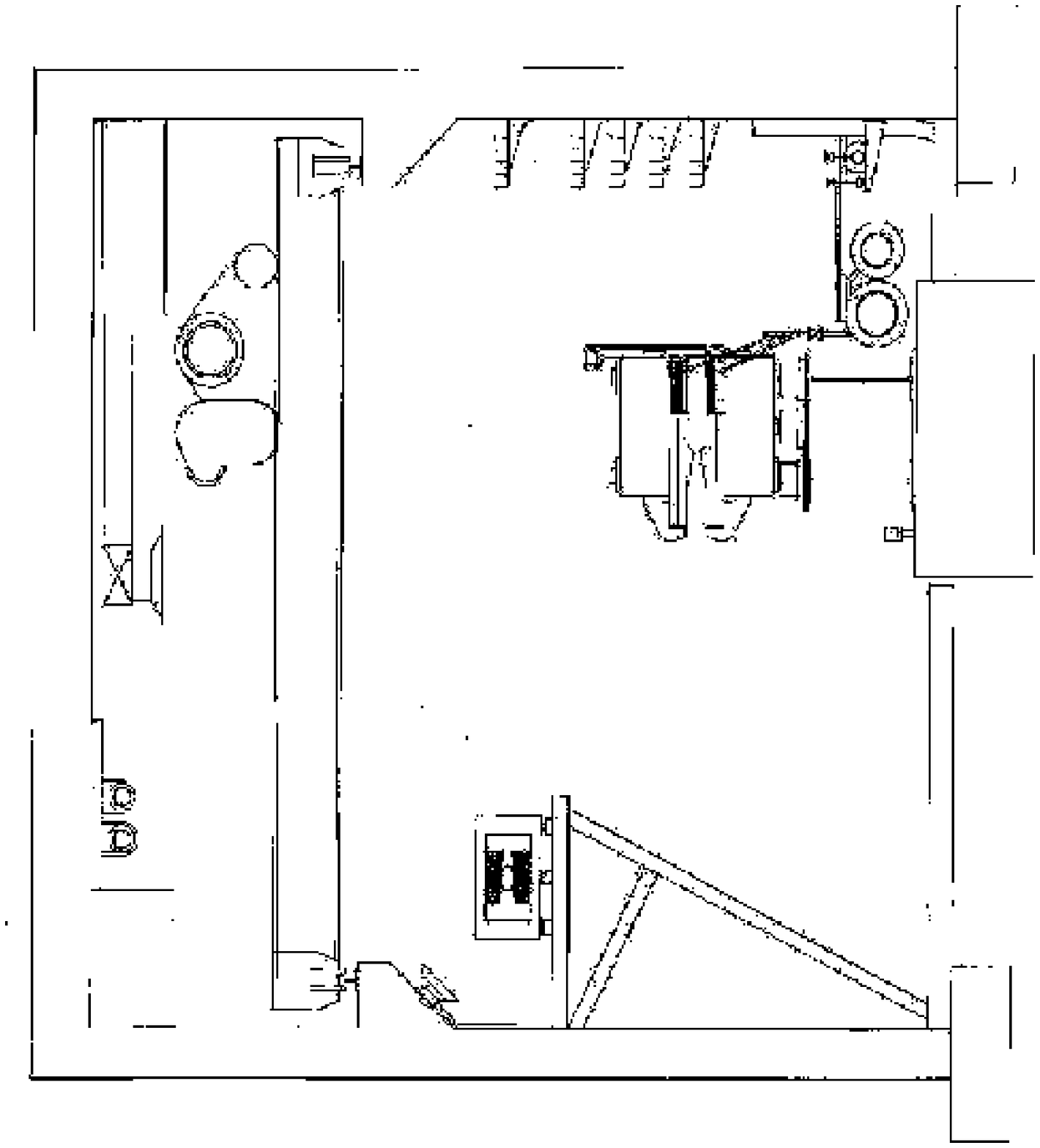}
    \caption{Placement of the permanent magnet accumulator ring inside the AGS tunnel.} 
    \label{page9}
  \end{center}
\end{figure}

\begin{figure}
  \begin{center}
    \includegraphics*[width=0.8\textwidth]{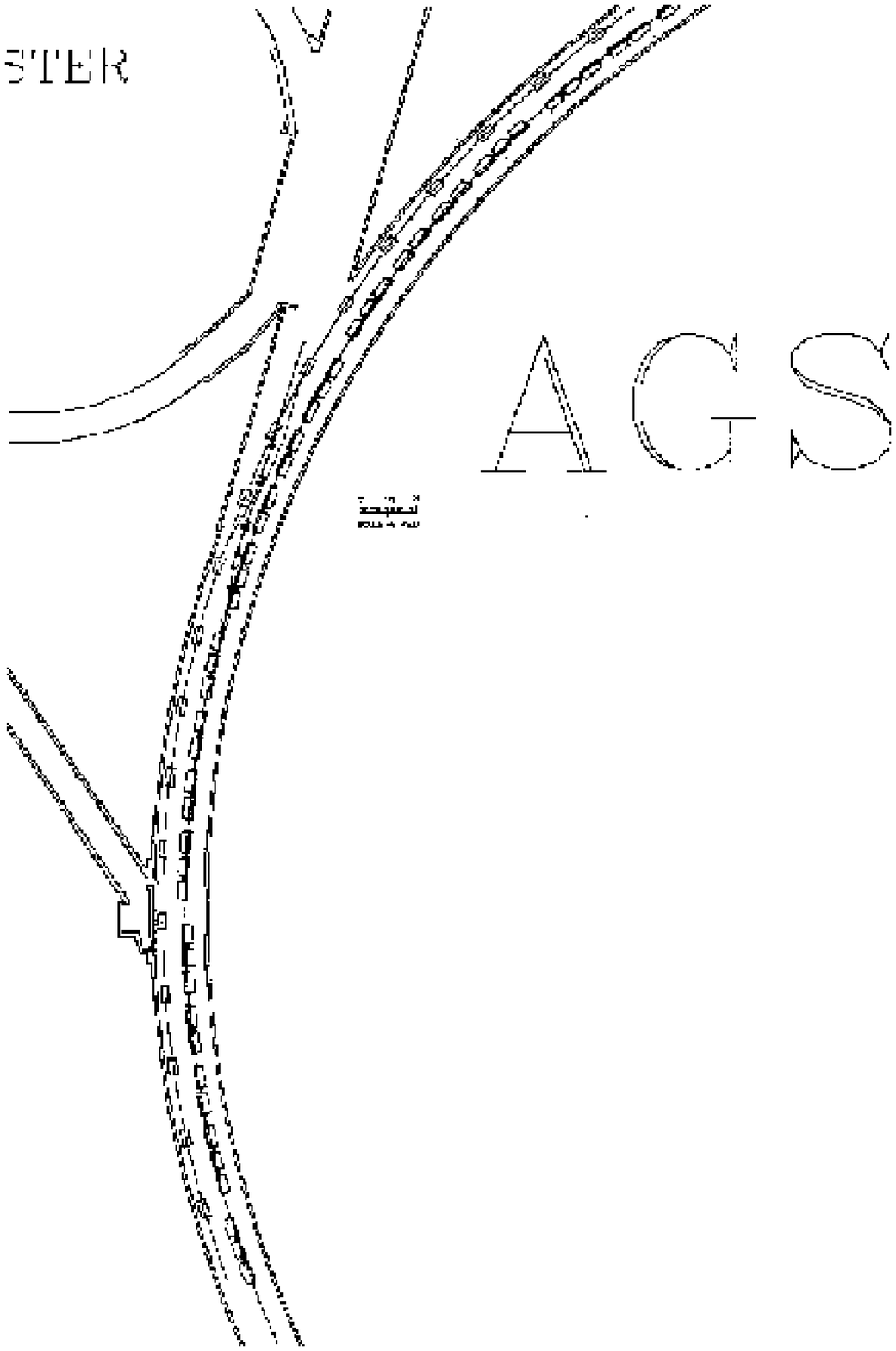}
    \caption{Placement of the permanent magnet accumulator ring inside the AGS tunnel. Satellite
      view.} 
    \label{page10}
  \end{center}
\end{figure}

\begin{figure}
  \begin{center}
    \includegraphics*[width=\textwidth]{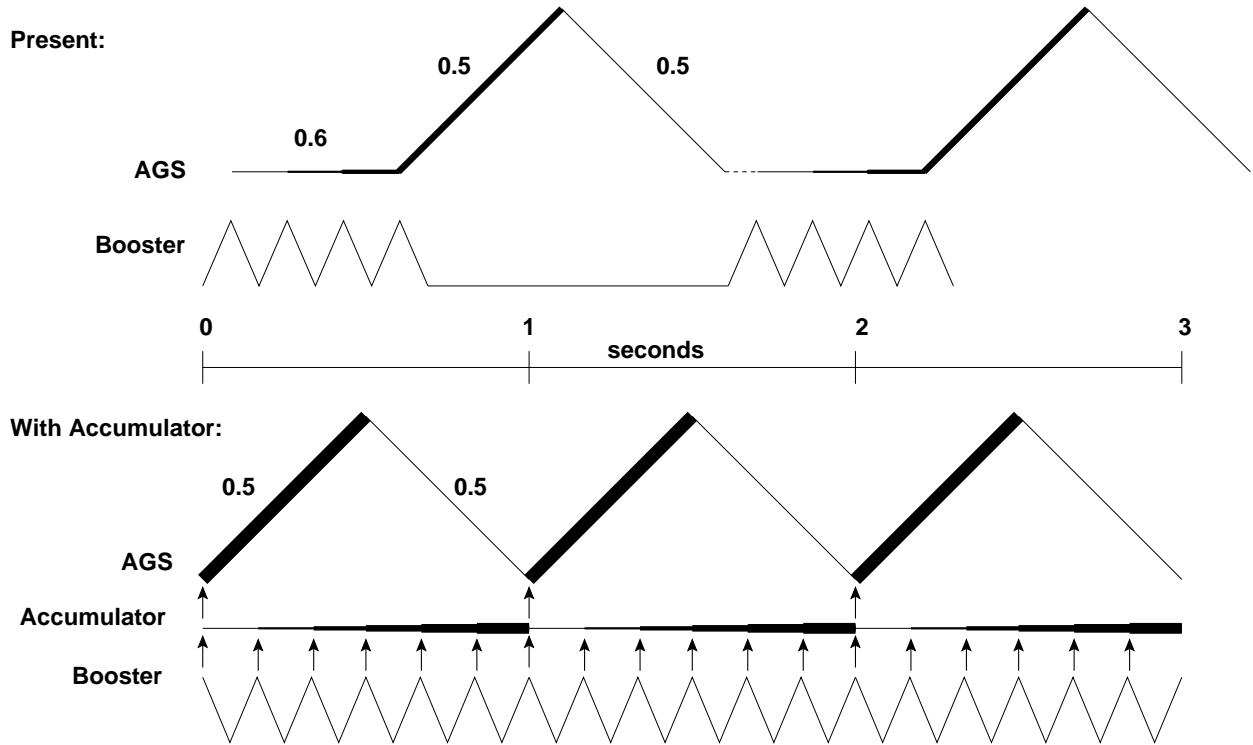}
    \caption{Time sequence of injecting pulses into the AGS. Top picture shows that
      at the moment 4 booster pulses are injected into the AGS during
      the period when the AGS magnets are at low field. In the new
      proposed configuration in the bottom picture the booster will
      inject 6 pulses into the accumulator which will store the beam
      until the AGS is at low field and then transfer the beam into
      the AGS.}
    \label{page4}
  \end{center}
\end{figure}

The preliminary design of the AGS upgrades and the new neutrino beam
has been produced by the AGS department to reach an AGS power of 0.53
MW in its first phase and 1.3 MW in the second phase \cite{roser01}. 
 In the first  phase the LINAC will be improved to
inject protons to the booster at 400 \MeV{} (at present it is 200
\MeV{}), and the booster energy increased to 2.5 \GeV{} from 1.8
\GeV{}. The addition of a fixed field accumulator storage ring between
the booster and the AGS main ring will increase the AGS input beam
from the present 4 booster pulses per AGS acceleration to 6 booster
pulses per AGS acceleration and, at the same time, increase the AGS
frequency from 0.6 Hz to 1.0 Hz. The AGS power increase would be from
0.14 to 0.53 MW.  Figures~\ref{page3} to \ref{page10} show the
proposed additions.  The new accumulator will be in the same tunnel as
the AGS.  Figure \ref{page4} shows the present and proposed AGS
injection modes.  The AGS intensity upgrades and parameters are shown
in Table~\ref{agsupg}.  The location of the accumulator ring in the
AGS tunnel is shown in Figs.~\ref{page9} and \ref{page10} and the
accumulator parameters given in Table~\ref{page8}.  The fixed field
magnets of the accumulator ring will be essentially copies of the
magnets produced for a similar purpose at Fermilab \cite{foster}.  
In the second phase of the upgrades the AGS repetition rate will 
be increased to 2.5 Hz to reach a total beam power of 1.3 MW.

The tentative cost estimates for the various upgrades and improvements
are shown in Table \ref{agscost}, separated into two relatively
independent phases of roughly equal cost.  It is possible to rearrange
the order in which the improvements are made.  A partial study of the
options is shown in Table~\ref{agsupg}.  Three options are shown: 
\begin{enumerate}
\item only the LINAC is improved,
\item only the accumulator is added,
\item both are done.
\end{enumerate}
Some AGS and booster power supply upgrades are assumed along with the
accumulator.  A detailed study of space charge, intensity, injection
energies, rep rate, power and cost needs to be made to make a much
more specific proposal for this upgrade.  This would come about with
approval of the present letter of intent.

\begin{table}

\begin{tabular}{|l|l|l|l|l|l|}
\hline 
  &  Now  & 400 \MeV{}  &  \multicolumn{2}{c|}{2.5 \GeV{} Accumulator} & AGS to  \\
  &       &  LINAC  & 200 \MeV{}   &  400 \MeV{}                             & 2.5 Hz \\
  &       &         &  LINAC    &  LINAC         &                              \\
\hline 
LINAC Energy (\MeV{}) &  200 & 400 & 200 & 400 & 400  \\
Booster Intensity (ppp) & $1.5\times 10^{13}$ & $2.0\times 10^{13}$ & $1.5\times 10^{13}$ & $2.0\times 10^{13}$ & $2.0\times 10^{13}$ \\   
Booster energy (\GeV{}) & 1.8 & 1.8 & 2.5 & 2.5 & 2.5  \\
Booster Cycles & 4 & 4 & 6 & 6 & 6  \\
AGS energy (\GeV{}) & 24 & 28 & 28 & 28 & 28   \\
AGS Intensity (Tp/sec) & 36 & 48 & 90 & 120 & 300 \\
AGS Rep Rate (Hz) & 0.6 & 0.6 & 1.0 & 1.0 & 2.5 \\
AGS Current ($\mu$ A) & 5.75 & 7.7 & 14.4 & 19.2 & 48 \\
AGS Intensity (ppp) & $6\times 10^{13}$ & $8\times 10^{13}$ & $9\times 10^{13}$ & $12\times 10^{13}$ & $12\times 10^{13}$  \\ 
AGS power (kW) & 138 & 215 & 403 & 538 & 1344 \\ 
\hline 
\end{tabular}
\caption{AGS Beam Power  Upgrade Plan.}
\label{agsupg}
\end{table}

\begin{table}
  \begin{center}
\begin{tabular}{ll}
Max. Kinetic Beam Energy & 2.5 \GeV{}  \\
Rigidity  & 13 Tm \\
Circumference & 819.07 m \\
Number of Superperiods & 25 \\
Number of FODO Cells & 50  \\
Number of (Combined-Function) Dipoles & 100  \\
Dipoles Length/Field & 3.25 m/ 0.25 T  \\
Length of FODO Cell & 16.45 m  \\
Length of straight sections & 8.22 m \\
Phase Advance per FODO Cell & 72 degrees  \\
  &  \\
Vacuum Chamber x/y & 150 mm/ 75 mm \\
Acceptance Geometric x/y & 234/58 $\pi$ mm mrad \\
Acceptance Normalized \@ 2.5 \GeV{} x/y & 744/184 $\pi$ mm mrad  \\
Space Charge limit:  &  \\
(100 $\pi$ mm mrad, $\delta \nu = 0.3$, 2.5 \GeV{}, debunched) & $3.5\times 10^{14}$ ppp \\
 &  \\
Betatron Tunes x/y  & 10.3/10.5  \\
Chromaticities x/y & -11.7/-12.7  \\
Beta Max. x/y & 24.7 m/24.3 m \\
Beta Min. x/y & 7.8m/7.5 m \\
Dispersion Max. & 1.8 m \\
Transition Gamma & 9.6 \\

\end{tabular}
\caption{Parameters of the permanent magnet accumulator ring.}
\label{page8} 
  
  \end{center}
\end{table}

\begin{table}
  \begin{center}
\begin{tabular}{|l|r|}
\hline 
Phase I (AGS at 1 Hz) &  \\
\hline 
300 \MeV{} SRF (116 \MeV{} to 400 \MeV{})  & \$35 M \\ 
2.5 \GeV{} AGS accumulator ring & \$25 M  \\
AGS Injection at 2.5 \GeV{} & \$ 5 M  \\
Total for Phase I 0.53 MW & \$65 M  \\
\hline 
Phase-II (AGS at 2.5 Hz) &  \\
\hline 
AGS power supply & \$32 M \\
AGS RF upgrade & \$8.6 M \\
Booster Power Supply & \$5.5 M \\
AGS Collimation and Shielding & 8.0 M \\
Total for Phase-II 1.3 MW & \$ 54.1 M  \\
\hline 
\end{tabular}
\caption{Cost of upgrading the AGS in two phases to 1 MW. The superconducting 
LINAC upgrade could be delayed to be after the accumulator. In this case phase 
I could deliver about 0.3 MW at a cost of \$30 M.  It is assumed that the target station shielding can be retrieved from existing resources.} 
\label{agscost}
  
  \end{center}
\end{table}

\section{Neutrino Beam Design} 

\begin{figure}
  \begin{center}
    \includegraphics*[width=0.9\textwidth]{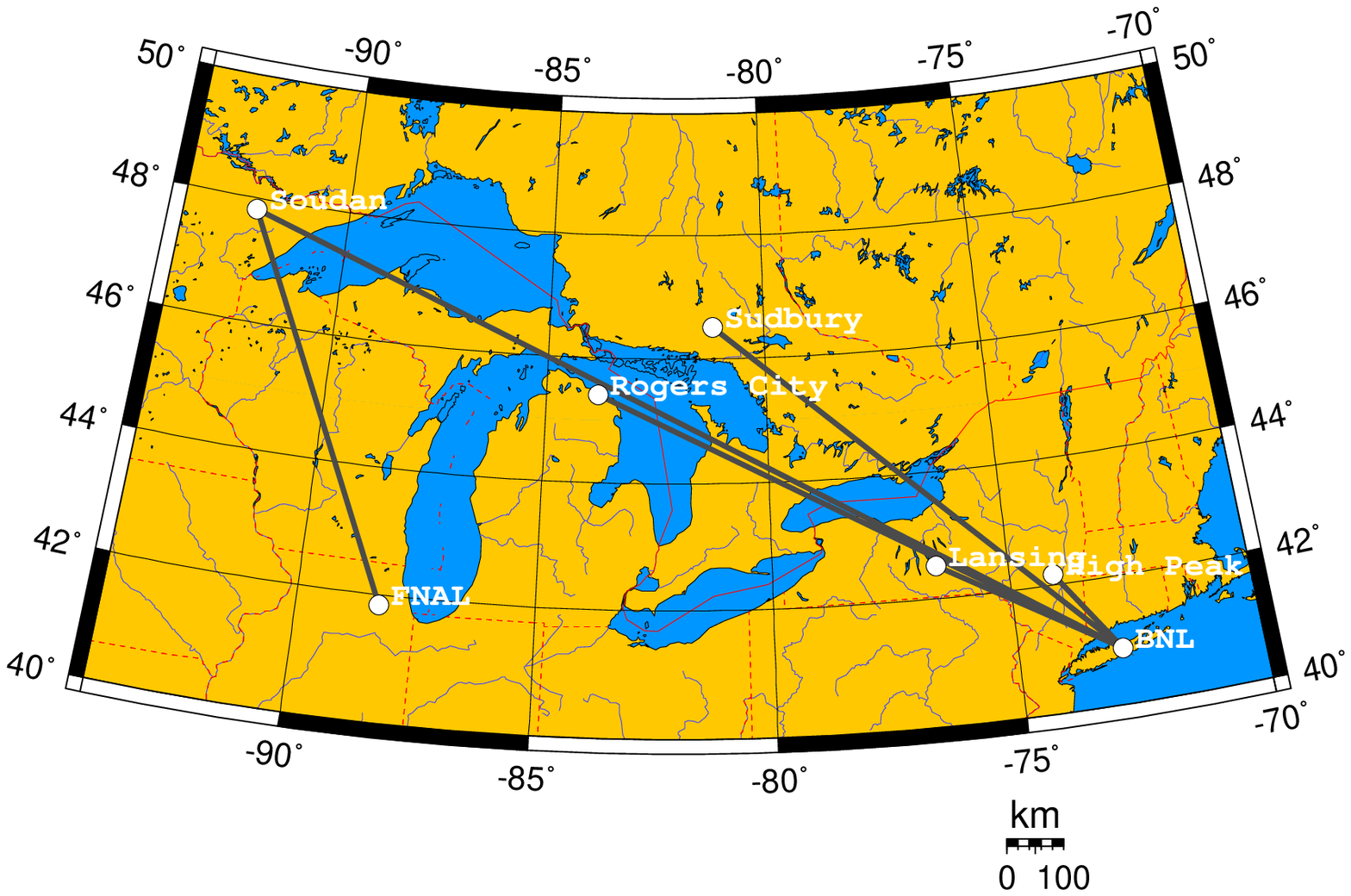}
    \includegraphics*[width=0.9\textwidth]{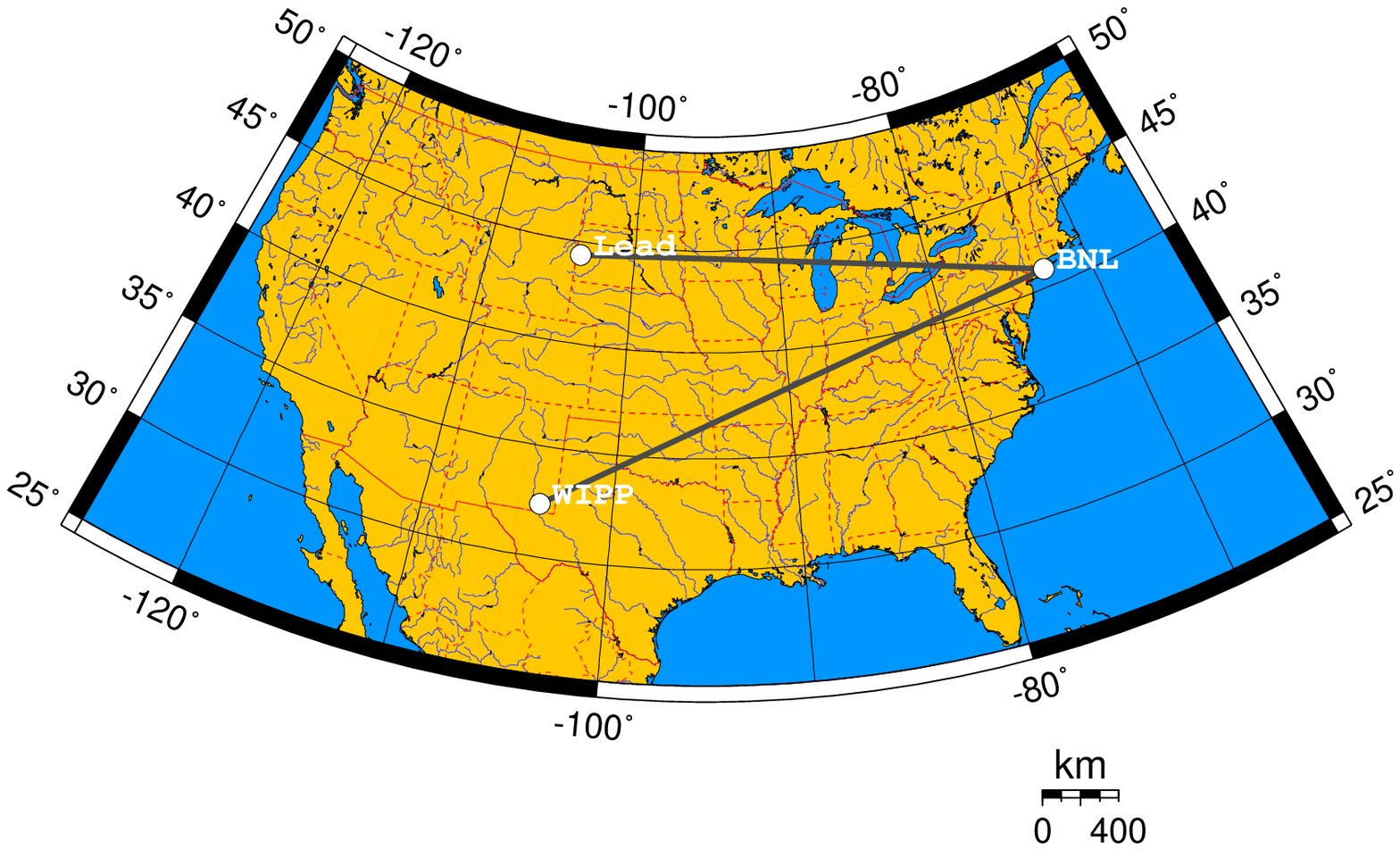}
    \caption{Possibilities for baselines from BNL.
      The distances from BNL to Lansing, Soudan, Lead (Homestake), and
      WIPP are 350, 1770, 2540, and 2880 km, respectively. }
    \label{blines}
  \end{center}
\end{figure}

The geographic location of BNL on one side of the continent allows us
to send beams to a variety of distances including very long baselines
of 2000 km or more.  This is shown in Fig. \ref{blines}.  The
distances from BNL to Lansing NY, Soudan MN, Lead SD(Homestake), and WIPP in NM
 are
350, 1770, 2540, and 2880 km, respectively.  The respective dip angles
are 1.7, 7.9, 11.5, and 13.0 degrees.  The difficulty of building the
beam and the cost increases with the dip angle.

\begin{figure}
  \begin{center}
    \includegraphics*[width=0.8\textwidth]{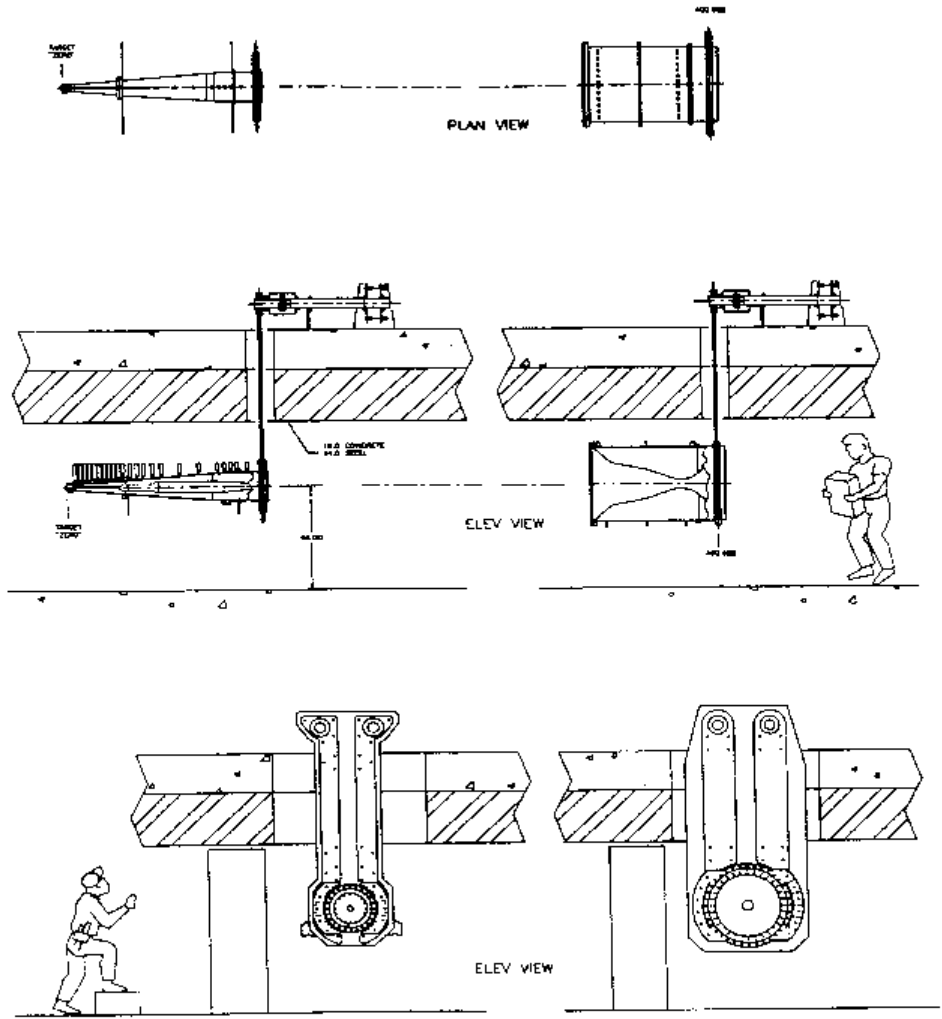}
    \caption{The design of the horn focusing system used for 
      the E734 experiment adapted from the E889 proposal.}
    \label{horns}
  \end{center}
\end{figure}

For the purposes of this LOI we use the design for a 
conventional wide band horn focussed neutrino beam similar 
to that   used in previous experiments at 
BNL such as  E734 (Fig.\ref{horns}).   
The design shown uses a water cooled copper target.
For much higher intensities this target will have to be 
redesigned. It is very likely that we can adapt the 
graphite target design used for NuMI at FNAL. 
We also need to modify the horn focussing to make a 
wider band beam to increase the neutrino flux in the 4 \GeV{}
region. There are a number of ways to optimize the horn
design; we will not discuss them here. 

\begin{sidewaysfigure}
  \begin{center}
    \includegraphics*[angle=-90,width=\textwidth]{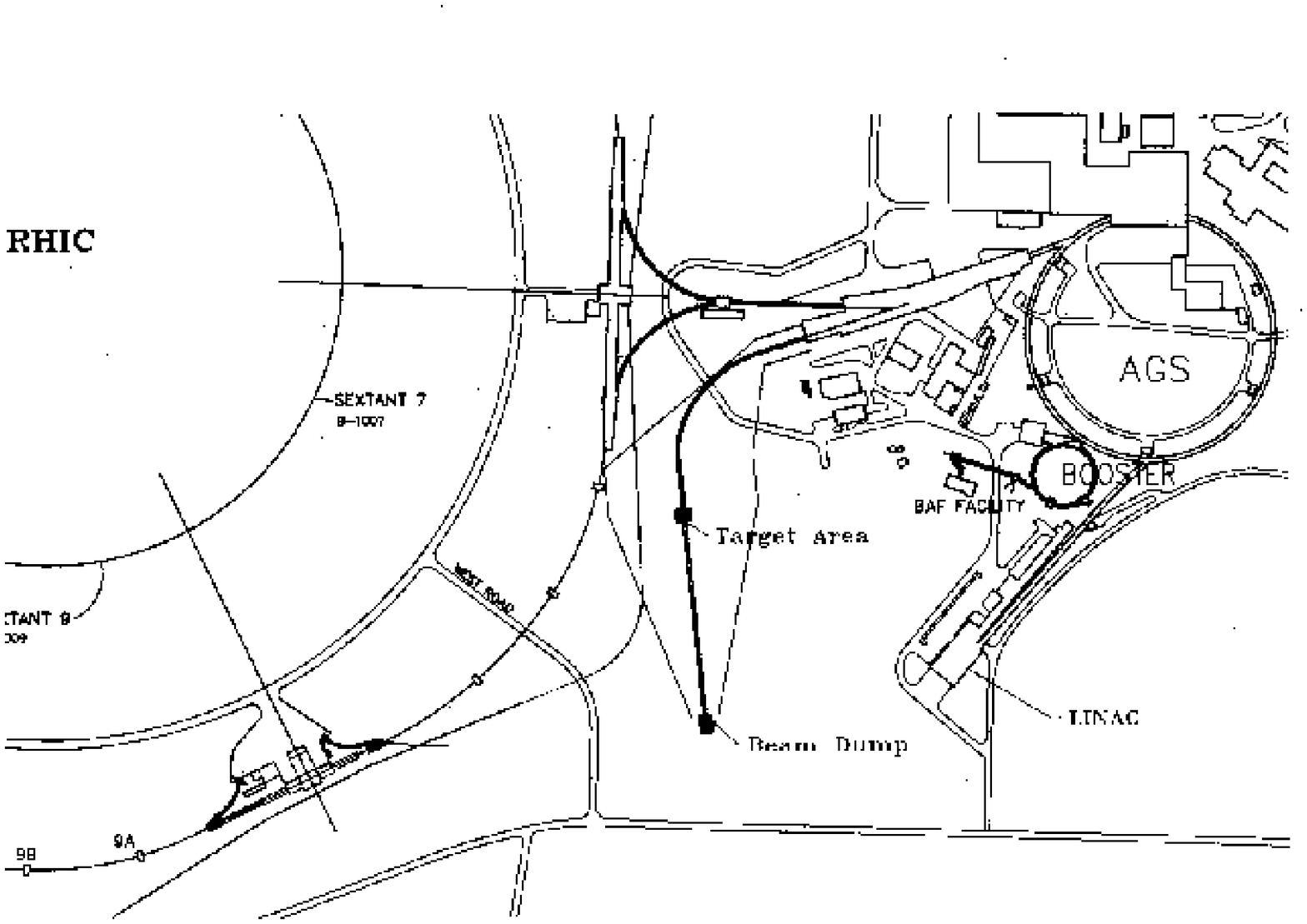}

    \caption{ The beam line for sending a neutrino beam to Homestake mine,
      South Dakota.
      This same beam line can be adapted for any far location in the Western 
      direction.}
    \label{planview}
  \end{center}
\end{sidewaysfigure}

\begin{sidewaysfigure}
  \begin{center}
    \includegraphics*[width=\textwidth]{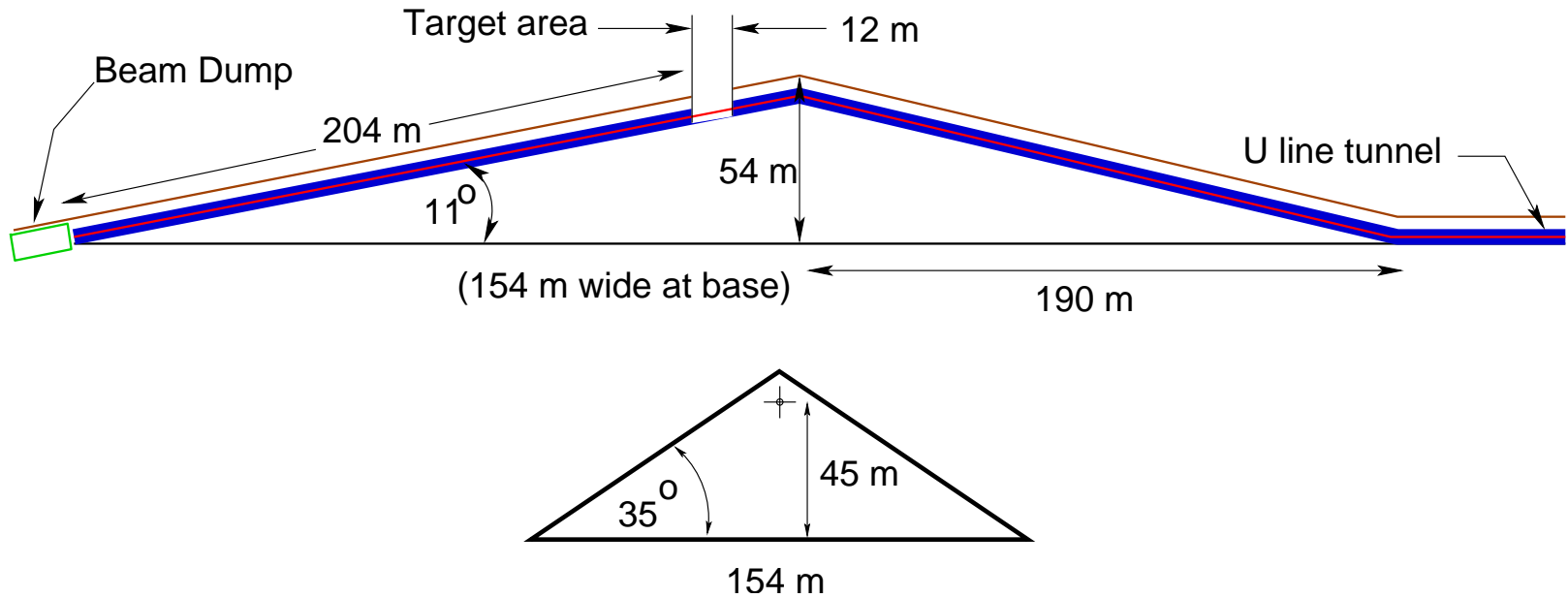}
    \caption{Elevation view of the neutrino beam line to 
      Homestake, South Dakota.  For a nearer location a much smaller
      hill can be constructed.  In this beam we assume a decay tunnel
      length of 200 m. For a shorter tunnel the cost of the hill will
      reduce as shown in table \ref{bcost}. }
    \label{eleview}
  \end{center}
\end{sidewaysfigure}

Our preliminary design for a beam to Homestake is 
shown in figures \ref{planview} and \ref{eleview}.
This can  be adapted to any far location in the western direction. 
Our design addresses a number of issues.
At BNL we are constrained to keep the beam line above the water 
table which is at a shallow depth ($\sim$ 20 m) 
  on Long Island. Therefore the beam has to be constructed on 
a hill that is built with the appropriate 11.5 degree slope. 
Fortunately, it is relatively easy, and inexpensive to 
build such hills on Long Island because of the flat, sandy 
geology. It is important to keep the height of the hill 
low so that the costs are not dominated by the 
construction of the hill. 
The proton beam must be elevated to
a target station on 
  top of the hill. The cost of the hill can be lowered  
by bending the proton beam upwards as quickly as possible.
We have, however, used the design and bend angle used 
for the RHIC injection lines for our design because 
the RHIC injection lines have well known costs. 

The new proposed  fast extracted proton beam line in the U-line tunnel will be a
spur off the line feeding RHIC. It will   turn almost due west, a few
hundred meters before the horn-target building. In addition to its 90
degree bend, the extracted proton beam will be bent upward through
13.76 degrees to strike the proton target.  The downward 11.30 degree
angle of the 667.8 ft meson decay region will then be aimed at the
2500 meter level of the Homestake Laboratory. This will require
the construction of a 39 meter hill to support the target-horn building,
so as to  avoid any penetration of the water table.  At its midpoint
(about Lake Michigan) the center of the neutrino beam will be roughly
120 km below the Earth's surface.

For a shorter baseline to Lansing NY in approximately the same direction 
as Homestake,   we would  not have to 
build the hill, which would lower  the cost 
by a considerable amount. We are considering a number of strategies 
for combining the proton transport and the target station 
for the two different baselines. 

A preliminary estimate of the cost without any of the customary 
burdens is shown in table \ref{bcost}. The costs are based 
on the the RHIC injector work, as well as the E889 proposal and 
the neutrino factory study.  
The conventional construction costs are 
dominated by the size of the hill which is 
approximately proportional to 
the third power of the decay tunnel length. In our cost
estimate we assume that we will bury the beam dump underground
to reduce the height of the hill.  
It is assumed that the target station shielding can be 
retrieved from existing resources. 
We have also estimated the 
cost assuming 200 m and 150 m long decay tunnels. 
The spectra shown in Fig. \ref{bnlspec} are based on a 200 m long tunnel. 
A shorter tunnel could reduce the intensity at higher energies 
because of the longer flight paths of parent pions. We will study 
this optimization further in future updates of this proposal.

\begin{table}
  \begin{center}
\begin{tabular}{|l|l|l|r|}
\hline
Item &  basis & 200 m & 150 m \\
\hline 
Proton transport & RHIC injector & \$11.85 M & \$11.85 M \\ 
Target/horn & E889 &  \$3.0 M & \$3.0 \\
Installation/Beam Dump & New & \$2.67 M & \$2.67 M \\
Decay Tunnel & E889 & \$0.45 M & \$0.45 M \\
Conventional const. (hill) & New & \$8.0 M & \$5.0 M \\  
Conventional const. (other) & E889 & \$9.1 M & \$9.1 M\\
\hline 
Total &      & \$35.19 & \$32.19 \\
\hline 
\end{tabular}
\caption{Preliminary cost of building the neutrino beam. The third column
is for a beam with 200 meter tunnel. The fourth column is for building the 
beam with a 150 meter tunnel.} 
\label{bcost}
  
  \end{center}
\end{table}

\vspace{1ex}
\section{ Conclusion}
\vspace{1ex}

We have outlined the neutrino physics program for an intense 
new neutrino beam from the Brookhaven AGS. 
The four goals of accelerator neutrino physics: precise determination of 
\dmatm{}, detection of \numunue{} appearance,
measurement of the matter effect, and detection of CP asymmetries in 
the neutrino section are all possible for the proposed complex for 
reasonable values of the oscillation and mixing parameters, some of which are not 
yet known.  
Further surprises in neutrino physics should not be discounted,
therefore any new facility must have sufficient flexibility to address 
new challenges. Our proposal allows such flexibility because of the 
possibility to mount both  very long (over 2500 km) and intermediate (400 km)  
baseline experiments with  beam intensity that can be increased in 
stages.

The AGS complex is unique because it can be upgraded simply by
increasing the repetition rate 
 of the machine.  This ability allows us the
flexibility to continuously upgrade the facility to as much as 2.5
MW~\cite{roser01}. In this proposal we have examined upgrades up to 1.3 MW.
The estimated cost of the first phase of the AGS upgrades to reach
0.53 MW, plus the new neutrino beam directed to Homestake is
approximately \$100M.  With a 30 percent contingency, the total cost is
\$130M.  It is probable that the first three modules of the detector
array will be produced in about five years, so that construction of
the AGS upgrades and neutrino beam would be planned for that period
and involve an average expenditure of approximately \$30M/yr.  For a
detector at intermediate baseline the costs will be less.  The total
yearly cost to the AGS department to provide protons for and maintain
the neutrino beam would be about \$9M, approximately equal to the
operations expense at present for HEP experiments.  Neither the
duration of the construction period nor the anticipated cost of the
improvements to the BNL AGS complex is large in relation to plans and
expenditures now usual for major apparatus in high energy and
elementary particle physics.
  Moreover, the
improvements to the AGS and the new beam line will be available for
carefully chosen other physics (for example, rare muon and kaon 
decays as well as muon EDM measurements) 
 in addition to providing important
advances in our understanding of this exciting new frontier of
elementary particle physics.


\begin{thebibliography}{99}



\bibitem{homestk}
National Underground Science Laboratory at Homestake, Lead, SD,  \\
http://mocha.phys.washington.edu/NUSL/


\bibitem{wipp}
Waste Isolation Pilot Plant, Carlsbad, NM, \\
http://www.wipp.carlsbad.nm,us/

\bibitem{sk}
S. Fukuda et al., Phys. Rev. Lett. {\bf 86} 5656, 2001;  
E.W. Beier, Phys. Lett. {\bf B283}, 446 (1992); 
T. Kajita and Y. Totsuka, Rev. Mod. Phys. {\bf 73}, 85 (2001).

\bibitem{imb}
C. McGrew {\it et al.}, Phys.\ Rev.\ D {\bf 59}, 052004 (1999).

\bibitem{sno} 
Q. R. Ahmad et al., Phys. Rev. Lett. {\bf 87} 071301 (2001). 
S. Fukuda et al., Phys. Rev. Lett., {\bf 86} 5651 (2001). 

\bibitem{lsnd}
C. Athanassopoulos et al., Phys. Rev. Lett. {\bf 77} 3082 (1996); 
 C. Athanassopoulos et al., Phys. Rev. Lett. {\bf 81} 1774 (1998) 

\bibitem{boon}
Booster Neutrino Experiment, Fermi National Laboratory, \\
http:/www-boone.fnal.gov/


\bibitem{k2k}
S. H. Ahn et al., Phys. Lett. {\bf B 511} 178 (2001).

\bibitem{j2k}
The JHF-Kamioka neutrino project, Y. Itow et al., 
arXiv:hep-ex/0106019, June 2001.

\bibitem{kasuga} S. Kasuga {\it et al.}, 
Phys Lett. {\bf B374}, 238 (1996).

\bibitem{minos}
Numi MINOS project at Fermi National Accelerator Laboratory, \\
http:/www-numi.fnal.gov/ 

\bibitem{cngs} 
CERN Neutrinos to Gran Sasso, \\
http://proj-cngs.web.cern.ch/proj-cngs/

\bibitem{e889}
 E889 Collaboration, Physics Design Report, BNL No. 52459, April, 1995. \\
http://minos.phy.bnl.gov/nwg/papers/E889/


\bibitem{e734}
L. A. Ahrens et al., Phys. Rev. {\bf D 34}, 75 (1986).

\bibitem{e734d}
L. A. Ahrens et al., Phys. Rev. {\bf D 41}, 3297 (1990).


\bibitem{arafune} Jiro Arafune, Masafumi Koike 
and Joe Sato, Phys. Rev {\bf D56}, 3093 (1997).

\bibitem{marciano} William J. Marciano, 
arXiv: hep-phy/0108181,  22 Aug 2001.

\bibitem{irina} Irina Mociouiu and
 Robert Schrock, 
arXiv: hep-ph/0106139v3, 15 Nov. 2001

\bibitem{wolfenstein}
By L. Wolfenstein (Carnegie Mellon U.). 1978.
In *West Lafayette 1978, 
Proceedings, Neutrinos '78*, West Lafayette 1978, 
C3-C6 and *Washington 1978, 
Proceedings, Long-distance Neutrino Detection*, 108-112.


\bibitem{study2}
 S.~Ozaki et al., eds.,
 {\sl Feasibility Study-II of a Muon-Based Neutrino Source}
 (June 14, 2001), 
 http://www.cap.bnl.gov/mumu/studyii/FS2-report.html


\bibitem{3m} 3M Collaboration: Proposal titled: Megaton 
Modular Multi-Purpose Neutrino Detector, Nov. 26, 2001.

\bibitem{uno} Physics Potential and Feasibility of UNO,
UNO collaboration, June 2001. 

\bibitem{lannddp}
 D.B.~Cline, F.~Sergiampietri, J.G.~Learned, K.T.~McDonald,
 {\sl LANNDD, A Massive Liquid Argon Detector for Proton Decay,
Supernova and
 Solar Neutrino Studies, and a Neutrino Factory Detector}
 (May 24, 2001), astro-ph/0105442 \\
Also see F.~Sergiampietri,
 {\sl On the Possibility to Extrapolate Liquid Argon Technology to a
Supermassive
 Detector for a Future Neutrino Factory},
 presented at NuFACT'01 (May 26, 2001),

\bibitem{larhighres}
F. Arneodo, et al., 
Nucl. Instrum. Meth. {\bf A461} 324 (2001)  


\bibitem{ICARUS}
F. Arneodo et al., 
 Nucl. Instrum. Meth. {\bf A471} 272-275 (2000) 


\bibitem{skahn}
M. Diwan, S. Kahn, R.B. Palmer (Brookhaven). Mar 1999.
Published in *New York 1999, Particle Accelerator, vol. 5* 3023-3025

\bibitem{argonprop}
 M.V.~Diwan et al., 
 {\sl Proposal to Measure the Efficiency of Electron Charge Sign
 Determination up to 10 GeV in a Magnetized Liquid Argon Detector
 ($\mu$LANNDD)},
 submitted to BNL (April 12, 2002), \hfill\break
 http://www.hep.princeton.edu/\~mcdonald/nufact/bnl\_loi/argonprop.pdf


\bibitem{roser01}
 M.J.~Brennan et al., ,
 {\sl 1 MW AGS proton driver},
 presented by T.~Roser at Snowmass'01 (June 2001), \hfill\break


\bibitem{foster} We have consulted
Bill Foster at Fermilab to design and calculate the cost of the
accumulator ring proposed here.

\bibitem{edm}
Muon Electric Dipole Moment experiment. \\
http://www.bnl.gov/edm/

\bibitem{rsvp}
Rare Symmetry Violating Processes, \\
http://meco.ps.uci.edu/RSVP.html


\end{thebibliography}
\end{document}